\documentclass{article}

\usepackage{arxiv}
\usepackage[flushleft]{threeparttable}

\usepackage[utf8]{inputenc} 
\usepackage[T1]{fontenc}    
\usepackage{hyperref}       
\usepackage{url}            
\usepackage{booktabs}       
\usepackage{amsfonts}       
\usepackage{nicefrac}       
\usepackage{microtype}      
\usepackage{cleveref}       
\usepackage{lipsum}         
\usepackage{graphicx}
\usepackage[round]{natbib}
\usepackage{doi}

\title{Insurgency as Complex Network: \\
Image Co-Appearance and Hierarchy in the PKK}

\author{ \href{https://orcid.org/0000-0002-9092-2775}{\includegraphics[scale=0.06]{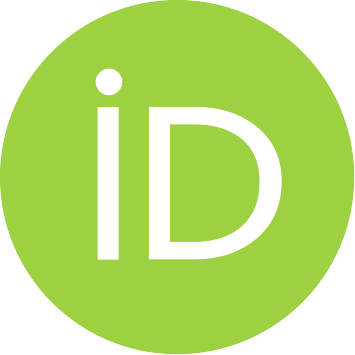}\hspace{1mm}Ollie Ballinger}\thanks{} \\
	Department for International Development\\
	Oxford University\\
	\texttt{ollie.ballinger@qeh.ox.ac.uk} \\
}

\hypersetup{
pdftitle={Insurgency as Complex Network},
pdfauthor={Ollie ~Ballinger},
}

\begin{document}

\twocolumn[ 
  \begin{@twocolumnfalse} 
  
\maketitle

\begin{abstract}
Despite a growing recognition of the importance of insurgent group structure on conflict outcomes, there is very little empirical research thereon. Though this problem is rooted in the inaccessibility of data on militant group structure, insurgents frequently publish large volumes of image data on the internet. In this paper, I develop a new methodology that leverages this abundant but underutilized source of data by automating the creation of a social network graph based on co-appearance in photographs using deep learning. Using a trove of 19,115 obituary images published online by the PKK, a Kurdish militant group in Turkey, I demonstrate that an individual's centrality in the resulting co-appearance network is closely correlated with their rank in the insurgent group. 
\end{abstract}
\vspace{0.35cm}

  \end{@twocolumnfalse} 
] 


\section{Introduction}

The organizational structure of a militant group governs many of its fundamental characteristics including its robustness to different counterinsurgency strategies, ability to pursue negotiated settlements, and factionalization. Though the most widely cited scholarly analyses of conflict largely treat insurgent groups as unitary actors \citep{fearon_ethnicity_2003, collier_greed_2004, hoeffler_beyond_2013, cheibub_democracy_2009, blattman_civil_2010}, a growing body of literature seeks to understand the complex social and political organization of these groups through the application of concepts and methods from social network analysis (SNA). The main impediment to this type of research is data availability: in a review of the literature on SNA in the study of insurgency, \cite[p.~231]{zech_social_2016} conclude that “The covert nature of militant groups makes them a difficult subject to study in general, a problem that becomes particularly acute when the objective is to map out the very internal structure that militants go to great lengths to conceal.” In this paper, I introduce a \href{https://face-network.readthedocs.io/en/latest/?badge=latest}{new methodology} that leverages deep learning and unsupervised clustering to automatically generate a social network graph based on co-appearance in photographs-- an abundant but underutilized source of data often generated by militants themselves. I then demonstrate a close relationship between the topology of the resulting graph and the organizational structure of the insurgent group, whereby high-ranking militants are more central in the image co-appearance network. 

Though insurgent groups generally produce very little public data, there is one main exception; online visual propaganda is the cornerstone of most militant groups’ recruitment strategies, with content ranging from combat footage, to magazines, to social media posts \citep{baele_lethal_2020, dauber_visual_2014}. Over a five month period in 2007, U.S. forces in Iraq raided eight Al-Qaeda media labs and seized 23 terabytes of images and videos destined to be uploaded to the internet \citep{dauber_youtube_2009}. More recently, militants have taken to directly posting these images on social media themselves \citep{klausen_tweeting_2015}; even loosely knit movements such as the participants in the Capitol Hill riots often document their activities on platforms such as Parler, generating over 70 terabytes of image and video data. Photographs-- particularly those in which individuals appear with each other-- are a rich source of relational data and are plentiful in the context of most insurgencies. Though image co-appearances have been used to construct social networks of university students and wedding-goers \citep{lewis_tastes_2008, berry_friends_2006}, they have never been used to construct the social networks of militants. Thus, I seek to answer the following question: To what extent can co-appearance in militant photographs be used to understand the social, political, and organizational forces that structure an insurgent group?

The following analysis uses as its only input \href{https://oballinger.github.io/PICAN-data/}{20,000 publicly available images published online} by the Kurdistan Workers’ Party (Partiya Karkerên Kurdistanê, henceforth PKK), an insurgent group primarily active in Turkey. This unstructured image data is converted into a social network graph in three steps: faces are automatically extracted from images using deep learning, nodes are generated by identifying individuals across images via unsupervised clustering, and edges are formed by linking individuals who appear together in photographs. Because “posing in a photograph with others is a deliberate act, and is generally indicative of some social connection” \citep[p. 44]{golder_measuring_2008}, this process effectively measures the degree and nature of an individual’s social embedding within the rebel group. The result is the PKK Image Co-Appearance Network (PICAN), a social network graph that mirrors many known characteristics of PKK. The PICAN has a number of useful analytical properties including a strong relationship between node centrality and rank, as well as a temporal dimension extracted from image metadata. These enable the identification of political factions within the PKK, a better understanding of the group’s resilience to counterinsurgency tactics, and even insights on gender segregation among militants. Ultimately, this analysis demonstrates that the abundance of unstructured image data generated by insurgents can be harnessed to generate social network graphs that deepen our understanding of the organizational structure of militant groups. 

This paper proceeds as follows. Section \ref{sec:litreview} reviews the literature on the application of network analytic methods to insurgent groups. Section \ref{sec:data} describes the image dataset derived from online PKK obituaries and discusses the performative nature of the photographs therein and their implications for the resulting co-appearance network. Section \ref{sec:methods} outlines the three step methodological approach: extracting faces from images, clustering individuals’ faces across images, and generating a network based on co-appearance. The rest of the paper analyzes the properties of the resulting network: Section \ref{sec:qual} shows that a qualitative interpretation of different measures of centrality enables a rudimentary distinction between members of the political and military wings of the PKK, and even captures the marginalization of key figures following failed leadership struggles. Section \ref{sec:rank} demonstrates a robust relationship between node centrality and rank within the PKK by cross-referencing nodes in the PICAN with wanted persons lists maintained by the Turkish government. Section \ref{sec:robustness} analyzes the relationship between network structure and the outcomes of various counterinsurgency tactics, with particular focus on the group’s regrowth following a military defeat. Section 9 concludes the paper.

\section{Literature Review}
\label{sec:litreview}

Though insurgent groups come in all shapes and sizes, the scholarly endeavor to understand the causes and conduct of civil conflict has largely ignored the internal characteristics of the actors involved. \cite{sanin_networks_2010} illustrate the importance of these organizational factors using two examples: the FARC-- a highly centralized "peoples' army" divided into hierarchically organized military units--and the Taliban, a loose confederation of semi-autonomous factions organized along tribal lines. Key outcomes in both of these conflicts are directly tied to group structure, including the feasibility of negotiated settlements; successive peace deals in Afghanistan failed in no small part due to the need to satisfy a diverse array of factions within the Taliban (such as the powerful Haqqani Network), as well as the individual proclivities of their leaders \citep{e_neumann_bringing_2014}. In contrast, the 2016 peace accord with the FARC was facilitated by the fact that "leadership had to convince its membership—something much easier in a highly hierarchical organization such as the FARC-EP" \citep{segura_made_2017}. 

The application of concepts and methods from Social Network Analysis (SNA) to the study of insurgent movements has generated a number of theoretical propositions related to the topology of militant networks. \cite{eilstrup-sangiovanni_assessing_2008} argue that decentralized networks such as Al-Qaeda and the Taliban are flexible and adaptable, but face challenges with internal cohesion and decision making that can make them prone to excessive risk-taking. \cite{stohl_networks_2007} contend that many ethnicity-based insurgencies such as Hamas, Hezbollah, and ETA form more centralized structures: small-world, scale-free networks which are characterized by short degrees of separation between nodes, well-connected hubs, and a degree distribution that follows a power law. These types of networks are particularly robust to the removal of nodes at random because only a small number of nodes are responsible for the overall connectivity of the network \citep{verma_economics_2020}. Due to difficulties inherent to the study of dark networks, “Empirical studies have mostly not engaged the central concerns of the theoretical literature—the relative advantages of centralized versus decentralized networks, the relationship between operational capability and network density, and the relevance of scale-free degree distributions” \citep[p. 231]{zech_social_2016}. In the remainder of this section, I highlight the ways in which the current approach can help address three key challenges evident in the empirical literature: data availability, selection bias and the use of secondary sources, and ambiguous relational ties.

\subsection{Data Availability}

The primary impediment to the application of SNA to the study of insurgent groups and other dark networks is that these organizations are often “purposefully attempting to remain opaque” \citep{morris_sna_2013}. Yet virtually every insurgent group maintains some form of online presence, and in most contexts their use of visual propaganda has eclipsed other media in their public relations \citep{dauber_youtube_2009}. Images have fast become one of the main sources of data for empirical analyses of individual insurgent groups; researchers have used the vast quantities of image data generated by insurgent groups to better understand their ideological foundations \citep{baele_lethal_2020}, shifts in their recruitment strategies \citep{lakomy_recruitment_2019}, and processes of collective identity formation \citep{macdonald_visual_2021}. At the same time, a number of studies have used image co-appearance to construct social networks in other contexts, recognizing that valuable information on social relations can be gleaned therefrom. However, these have either required manual coding \citep{berry_friends_2006, golder_measuring_2008}, or some level of structure in the image data such as the tagging of individuals in Facebook photos \citep{lewis_tastes_2008}. In the context of militant photographs, image data is almost certain to be unstructured and the manual recognition and tagging of thousands of unknown individuals is unfeasible.

By leveraging facial recognition technology and unsupervised classification to automate the generation of an image co-appearance network, the approach taken herein is far more scalable than manual coding, and can utilize completely unstructured image data. Both of these features are critical given that image data in the context of insurgent groups are likely to be both voluminous and unstructured, with basic information including the number of individuals contained in the photographs likely to be unknown. Furthermore, the image processing pipeline allows for the collection of a rich array of node-level attributes through the extraction of image metadata including location, date, camera type/serial number, as well as apparent age and gender estimation using deep learning. The code used to turn unstructured image data into a co-appearance graph has been published as an open source \href{https://face-network.readthedocs.io/en/latest/?badge=latest}{Python package}. Using three functions, researchers can easily extract faces from images, cluster faces based on similarity, and generate a graph based on co-appearance. The package features a flexible implementation that can be optimized according to the nature of the image data at hand; this includes options to extract metadata and perform apparent age and gender estimation, enabling attribute-based analysis (e.g. homophily, ERGMs, clustering) and where datetime metadata are available, analysis of the network’s evolution over time.

\subsection{Selection Bias and Secondary Sources}

Due to the extreme scarcity of primary source data generated by covert organizations, the vast majority of existing SNA studies of insurgent groups rely on secondary data, including newspaper articles \citep{rodriguez_march_2005, metternich_antigovernment_2013, krebs_mapping_2002}, legal documents \citep{jordan_strengths_2008}, and public statements \citep{magouirk_connecting_2008}. This is not only time consuming-- thereby limiting scalability and resulting in smaller sample sizes--it introduces a measure of subjectivity into the analysis. \cite[p. 567]{koschade_social_2007} social network analysis of the Jemaah Islamiyah cell responsible for the Bali bombings in 2002 contains only 17 nodes, connected on the vague basis of having had “numerous weak contacts over the period in question” or having resided together. This requires significant prior knowledge of the actors involved, limiting the novelty and utility of insights gleaned from the analysis. It also imposes an extreme degree of selection bias that effectively precludes the inclusion of anyone below the highest echelon. 

Selection bias is generally less severe in the rare studies that rely on primary data. Using a trove of thousands of ISIS job applications, \cite{edgerton_suicide_2022} is able to study suicide bomber mobilization and kin/peer ties at the level of individual recruits. Indeed, using ego networks constructed through interviews, \cite{stys_trust_2022} found that even complete data on "covert" group membership neglects important actors, as the liminal space between the rebel group and society is populated with individuals (e.g. demobilized combatants) who can act as brokers or fulfill other important roles. The use of photographs does not fully solve the issue of selection bias: some individuals may be systemically important but camera shy, while others may simply be photogenic. The extent of this bias depends on the context in which pictures are taken, and is discussed further in the following section. However, social networks created using image co-appearances are likely to yield a far more complete picture of a militant group than those created with secondary sources; lower level commanders and foot soldiers are unlikely to appear in news articles and legal documents, but may well appear in a group photograph, footage from a patrol, or pictures of a ceremony. Thus, a key advantage of the present approach is that it utilizes primary data in the form of images published by militant groups, and can yield a more complete picture of an organization.

\subsection{Ambiguous Relational Ties}

A significant methodological concern in the application of network analytic methods to the study of militant groups is “the use of implicit, poorly characterized relational ties in many studies. Some analyses use relational data coded to cover a wide range of social exchange” \citep[p. 231]{zech_social_2016}. In a study of the Provisional Irish Republican Army (PIRA), \cite{gill_lethal_2014} create a network of relational ties between individuals including marriage, friendship, blood relation, and co-involvement in PIRA activities. Even within relational ties of the same type (friendship, kinship, cohabitation), there can be substantial variation in the weight of the tie that is rarely accounted for \citep{pedahzur_changing_2006}. If the ties between nodes encompass a diverse range of social interactions, network statistics and hypothesis tests derived from an aggregate quantitative analysis of these edges lose conceptual clarity. 

Co-appearances in images have been used to construct social networks in several influential studies, and are generally recognized as signalling significant social ties between individuals. \cite{berry_friends_2006} argues that the use of co-appearances in wedding photographs provides a less subjective measure of interracial friendships than traditional measures thereof including surveys. In the context of facebook photos, \cite{lewis_tastes_2008} suggest that co-appearance reflects a relatively high level of positive affect between the individuals involved, as well as a desire to be socially recognized with the individual. Though the process underlying when and where militants take pictures is governed by specific social practices (e.g. training academy graduations, platoon photographs, celebrations), co-appearance acts as a reliable proxy for the existence of personal contact between individuals. This, in turn, plausibly measures the extent of an individual’s social embedding within an organization: if the same person is persistently present in group photographs in a militant organization, there is likely to be a reason. 

\begin{figure*}[!h]
  \caption{A Photoshoot with the PKK's Leader}
  \begin{minipage}{\textwidth}
  \includegraphics[width=\textwidth]{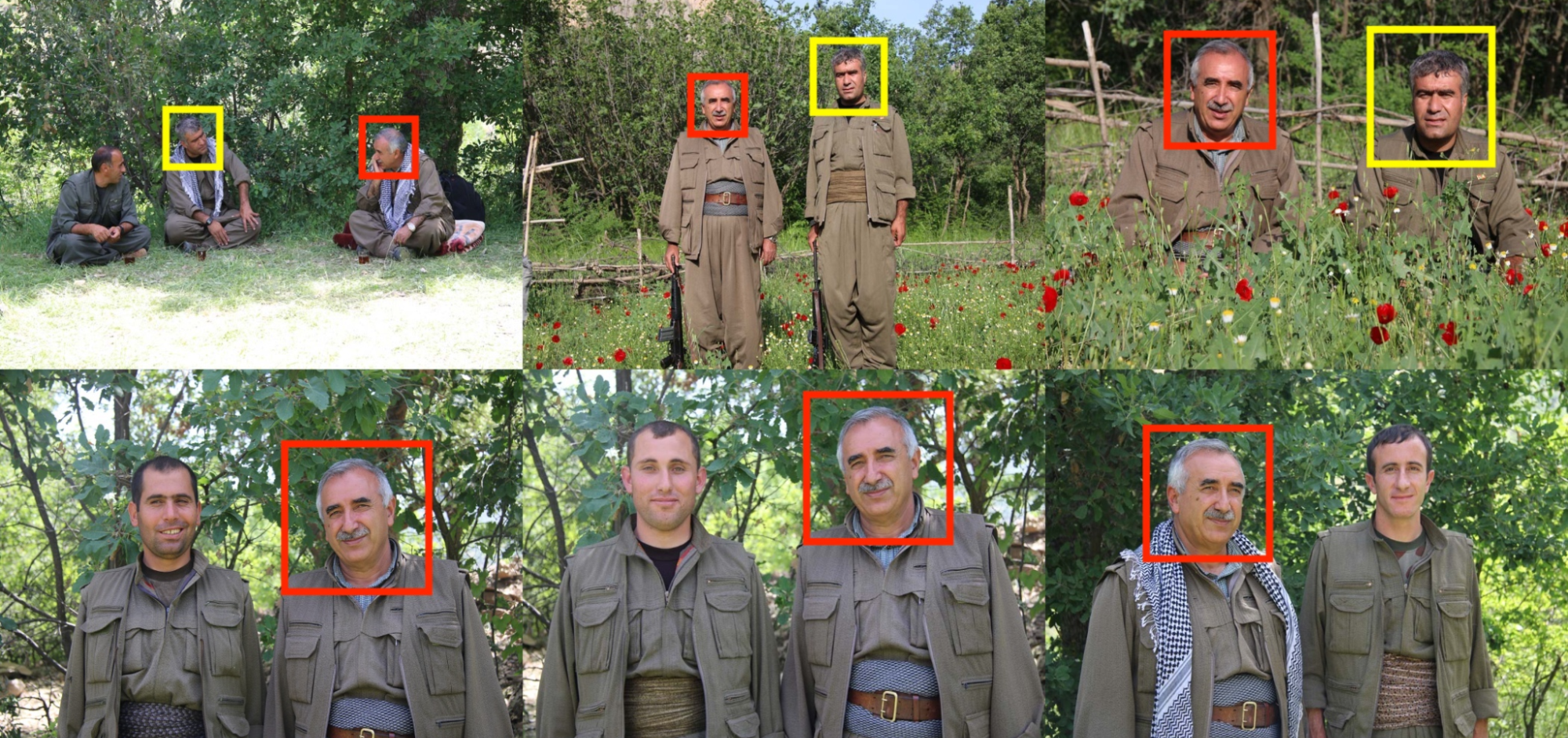}
{\scriptsize These images were taken between 07:24 and 12:41 on 05/05/2014, using a Canon EOS 70D with the serial number “43021010637”. They depict Murat Karayilan, the current leader of the PKK, indicated by the red box. They also feature Ekrem Güney, a mid-level commander, indicated in yellow.\par}
\end{minipage}
    \label{figure1}
\end{figure*}

\section{Data: Militant Photographs}
\label{sec:data}

When a member of the PKK is killed, an entry for the individual is added to the group's \href{https://hpgsehit.com}{obituary website}. Each obituary features photographs of the recruit, including a high definition portrait. These photographs appear to have been taken for a wide variety of reasons. The most common type of photograph is a posed portrait of the fallen militant. Others feature militants posing next to a commander. Many appear to be taken at various ceremonies and celebrations including funerals, training academy graduations, or during Nowruz. Photographs are not taken during executions of suspected traitors, nor are those who suffer this fate given the honor of an obituary. As such, this image dataset represents a collection of the moments that PKK members and photographers have decided to capture, and the social interactions between the individuals therein. This section explores different aspects of this performance and how they are likely to manifest themselves in a co-appearance network. 

\subsection{Posed Photographs}

19,115 images were downloaded from 2484 individual obituaries dated between 2000 and 2021. An \href{https://oballinger.github.io/PICAN-data/}{interactive visualization} generated using the entire image dataset used herein allows readers to qualitatively explore the photographs. The selection of many of these moments and individuals are doubtlessly curated to present an “idealized front” which furthers particular narratives about the PKK. For example, the representation of women in these images is not only a manifestation of the PKK’s feminist ideological tenets, but is an integral part of the group’s public relations and recruitment strategies. \cite{wood_female_2019} notes that “The image of a smiling young woman holding an assault rifle has become a common feature of media reports about the Partiya Karkerên Kurdistanê (PKK)”. Indeed, women are slightly over-represented in these images: on average, women appear in 6.1 images, while men appear in 5.2. A T-Test suggests that this gendered difference in mean photograph appearances is significant (t=5.18; p<0.001). Because women appear in more photographs than men, they are likely to appear more central in the co-appearance network.

Similarly, images of insurgent leaders are inherently performative. Though a rebel group may want to avoid exposing their command structure in photographs, images of the leaders amongst the rank and file rather than cowering in caves serve to cast them as both fearless yet humble. If this were the case, we could reasonably expect a relationship between an individual’s seniority and the number of co-appearances: a foot soldier may only appear with one or two commanders, but a commander would appear with many foot soldiers. However, the ways in which hierarchy is discernible from these images may not be consistent at all levels of the command structure; despite being a pivotal figure in the Cuban Revolution, Juan Almeida Bosque’s likeness is nowhere near as ubiquitous as that of Che Guevara. Individuals who are systemically important but not particularly salient could be under-emphasized in an image co-appearance network.

The six images in Figure \ref{table1} illustrate how the portrayal of leaders as fearless and humble generates a positive relationship between image co-appearance and rank. These images were taken between 07:24 and 12:41 PM on May 5th, 2014, using a high-end camera (Canon EOS 70D) with the serial number “43021010637”. They were the only images taken with this camera. 

An older man is present in all six pictures. In the top three photos he poses with the same man (who appears to be in his 50s), including one in which they are seated with a third man, sharing tea and talking. In the three remaining photos, he poses next to a different young man each time. The older man present in all six images is Murat Karayilan, the 67-year-old current leader of the PKK. The middle-aged man is Ekrem Güney, a mid-level commander who was killed in 2016. The three young men appear to be foot soldiers. A co-appearance network generated from these images would see Karayilan with seven edges, Güney with four, and the foot soldiers with one edge each. This type of posing links rank with co-appearance in such a way that it not only separates the leadership from the infantry, but correctly orders the leadership in terms of rank. Though these images appear to be from a photoshoot-- all taken with the same camera within hours of each other--this general principle reasonably extends to other social practices captured in these photographs in which posing is common, including ceremonies and holidays. 

\subsection{Candid Cameras}

Though many of these images are staged and curated to advance certain narratives of the PKK, the majority appear to be relatively candid. There are two likely reasons for this. The first is that paradoxically, candid pictures in this context make for better propaganda. In a study of online representations of martyrdom during the Arab Spring, \cite{halverson_mediated_2013} found that the depiction of martyrs as being relatable was crucial in generating the “imagined solidarities” that translate the collective identification with the martyr into support for a social movement. As an online repository of PKK obituaries, the emotive force that these images draw on is empathy-- they depict seemingly ordinary people eating meals together, playing backgammon, laughing; ordinary people who were killed by the Turkish state. Because empathy relies on one's ability to place oneself in an other's position, candid images in this context are far more affective than staged propaganda photos.

The second reason many of these images appear candid is that the PKK publishes more traditional and overt propaganda elsewhere: not only is this obituary website not the primary way in which the PKK disseminates propaganda to the wider world, it's not even the PKK's main website (hpgonline.com). The group also publishes several magazines (Serxwebun and Berxwedan), maintains social media accounts on various platforms, and even broadcasted an international satellite television channel (Roj TV) which has been replaced by an online streaming platform (Gerîla TV). The PKK disseminates a wide range of content on these platforms, ranging from combat footage to poetry. Thus, while the PKK makes extensive use of online visual propaganda to portray itself to a public audience, this is not the primary function of the obituary website from which the images used in this study are derived.

\begin{figure}[!h]
  \caption{Cameras Used by the PKK}
  \begin{minipage}{\columnwidth}
  \includegraphics[width=\columnwidth]{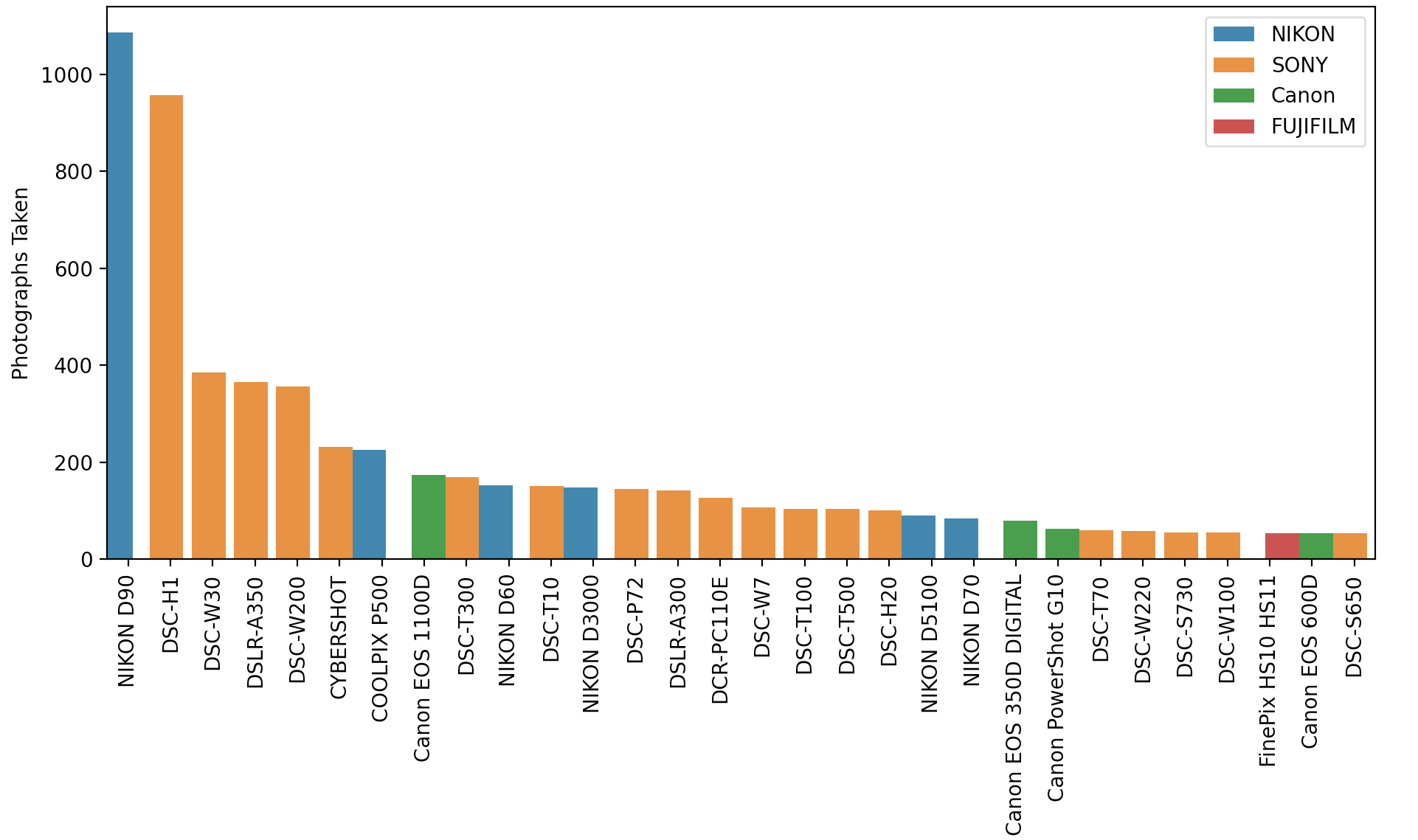}
{\scriptsize The Y axis shows the number of pictures taken with each type of camera. The plurality are taken with a high-end cameras, but the majority are taken with cheaper models.\par}
\end{minipage}
  \label{figure2}
\end{figure}

Metadata extracted from these photographs contains the make and model of the cameras employed by the PKK, supporting the proposition that many of these images were taken by amateurs. Figure \ref{figure2} shows the number of photographs in the sample taken with different types of cameras. Though the plurality of these images were taken using professional DSLR cameras such as the Nikon D90 or the Sony DSC-H1, the majority were taken with cheaper digital cameras accessible to amaterus such as the Sony DSC-W30 and W2000. The notion that many of the images taken with these cameras are “candid” is not meant to imply objectivity-- the deliberate presentation of candor is itself a form of performance. Rather, it simply emphasizes different actors and social relations: this style of photography likely increases the embeddedness of low-ranking individuals in the network by including group pictures of friends and comrades.

\section{Methodology}
\label{sec:methods}

\begin{figure}[h]
  \caption{Automatic Image Co-Appearance Network Creation}
    \begin{minipage}{\columnwidth}
  \includegraphics[width=\columnwidth]{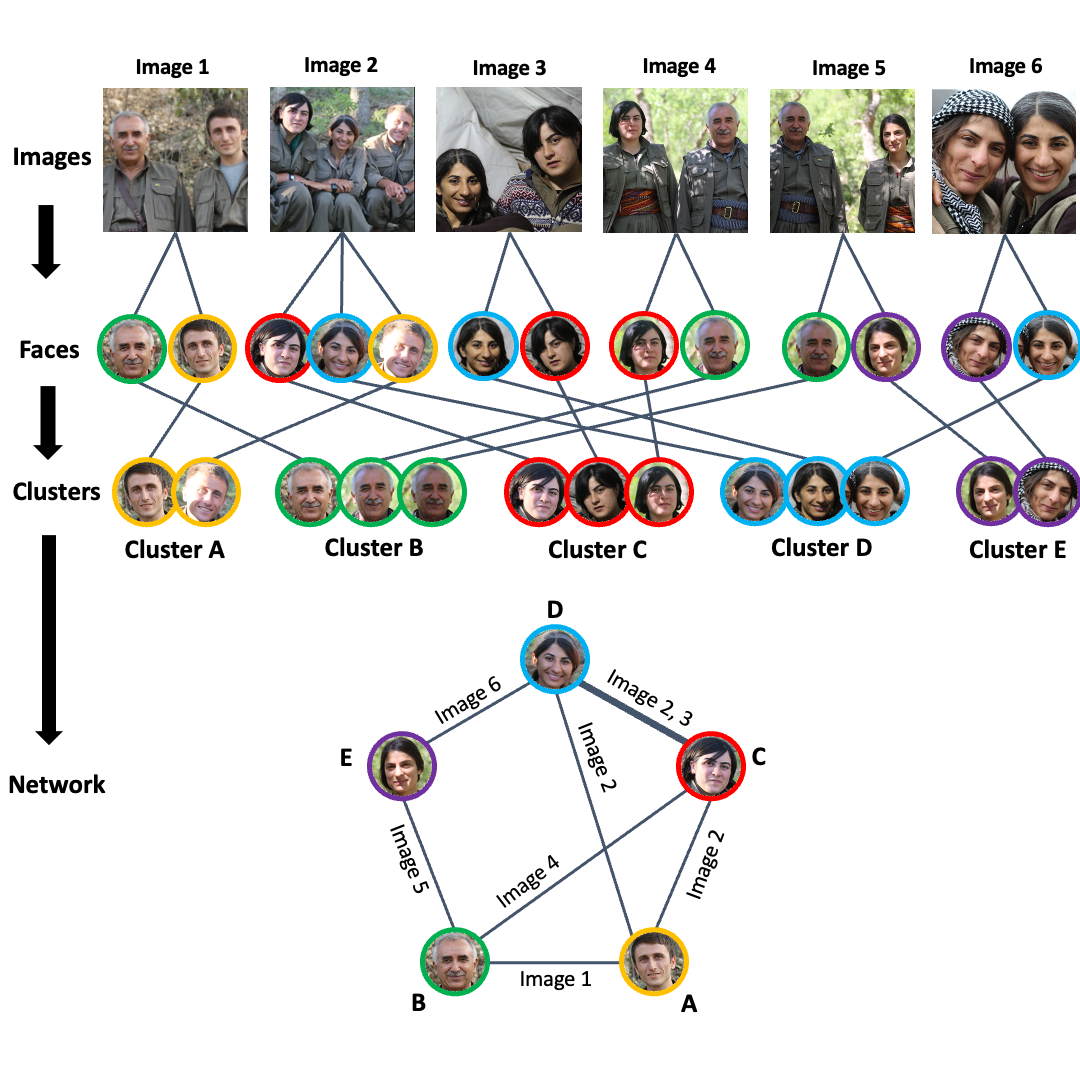}
{\scriptsize There are three steps in the automatic generation of an image co-appearance network; extracting faces from images, identifying the same face across images, and linking faces based on co-appearance\par}
\end{minipage}
    \label{figure3}
\end{figure}

Following collection, the unstructured image data is converted into a social network based on co-appearance in three main steps. First, I use a Residual Neural Network to extract facial encodings from images. I then use a graph clustering algorithm to identify individuals across images, generating a unique cluster of faces for each individual. Finally, I construct a weighted graph in which each cluster is a node, and each co-appearance constitutes an edge. These three steps are illustrated in Figure \ref{figure3}.

\subsection{Extracting Faces}

The first step in this analysis involves extracting faces from images. The approach taken herein relies on the FaceNet system developed by \cite{schroff_facenet_2015}, which converts face images to a “compact Euclidean space where distances directly correspond to a measure of face similarity”. These 128-dimensional face embeddings can be used to store, compare, and cluster faces based on similarity. A pre-trained Residual Neural Network with 29 convolutional layers was used to detect faces, generate embeddings, and perform apparent age and gender estimation. The model was accessed via the Python distribution of Dlib, which reaches an accuracy of 99.38\% on pairwise facial recognition using the standard “Labeled Faces in the Wild” (LFW) benchmark dataset \citep{king_dlib-ml_2009}.

In total, 23,496 faces were extracted from 19,115 images. Cropped face image tiles are generated to enable manual verification, and facial embeddings are stored for clustering. Inspection of face tiles reveals a number of interesting artifacts, including the extraction of artistic representations of Abdullah Öcalan in the form of flags, paintings, and even lapel pins; Samples of these clusters are provided in the appendix. These are discarded.

\subsection{Clustering Faces}

Once faces are extracted, they are clustered based on similarity to identify individuals across images. The accuracy of the clustering process is assessed through the construction of a labeled subset of the data. Though images from a given obituary will primarily contain the face of the deceased individual, they also frequently contain group photos in which other faces are present. However, if an image from an obituary only contains one face, this is almost always a portrait of the obituary’s subject. Thus, by restricting the sample of images to only those that contain one face, a labeled subset of the data is created. 

\begin{figure}
  \caption{Unsupervised Clustering Accuracy}
    \begin{minipage}{\columnwidth}
  \includegraphics[width=\columnwidth]{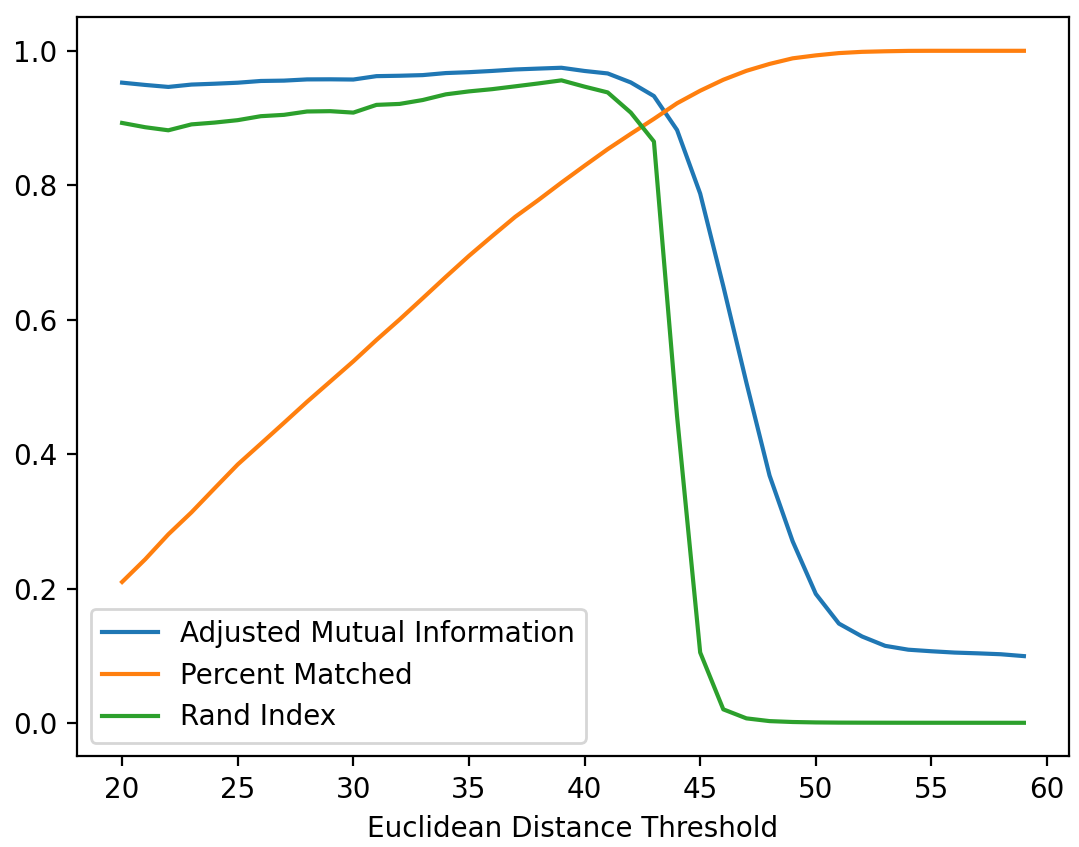}
{\scriptsize Effect on accuracy of tuning the euclidean distance parameter in graph clustering. The optimal threshold is 0.39\par}
\end{minipage}
  \label{figure5}
\end{figure}

The choice of the clustering algorithm is informed by a number of constraints in the data. Because some of the people who appear in the images are still alive, the number of unique individuals (and therefore, clusters) is unknown, precluding the use of algorithms such as k-means or spectral clustering. A number of unsupervised clustering approaches have been applied to the task of face clustering in recent years, including among others Agglomerative Hierarchical Clustering, Density-Based Spatial Clustering of Applications with Noise (DBSCAN), and Graph-Based Clustering. Graph-based clustering has recently been shown to achieve high accuracy in face clustering tasks \citep{chang_effective_2019}. This approach begins by creating an undirected weighted graph $G=(V, E)$ in which each vertex (V) is a facial embedding and each edge (E) is weighted according to the euclidean distance between vertices. In the construction of this input graph, only edges whose weights are below a certain euclidean distance cutoff are included. The optimal cutoff parameter is selected based on the results of tuning shown in Figure \ref{figure5}. 

Once the input graph is created, the Chinese Whispers (CW) algorithm is used to partition the nodes into clusters. CW functions by first assigning a unique class to each node. Through an iterative process, each node inherits the strongest class in its local neighborhood, which is the class with the highest sum of edge weights to the current node \citep{biemann_chinese_2006}. The CW algorithm achieves a remarkable degree of accuracy in the current dataset.

Maximum accuracy is achieved when the euclidean distance threshold is set to 0.39. If the cutoff is too low, false negatives increase as faces are not sorted into clusters. If the cutoff is too high, the number of false positives increases. The Rand Index, which measures the similarity between a set of labeled and predicted clusters, achieves a value 95.6\% under optimal conditions. The Adjusted Mutual Information index provides a similar assessment of clustering performance, but is more robust to unbalanced data with both very large and very small clusters; it achieves a value of 97.4\% under optimal conditions, with 80.3\% of face tiles sorted into clusters. The relatively high proportion of faces that are not sorted into clusters largely reflects poor image quality under certain conditions; large group photographs where some individuals are in the distance, low resolution images, or those taken with old cameras present a challenge to facial recognition. 

\begin{figure}[h]
  \caption{Auto-Generated Cluster for HPG Commander Bahoz Erdal}
      \begin{minipage}{\columnwidth}
  \includegraphics[width=\columnwidth]{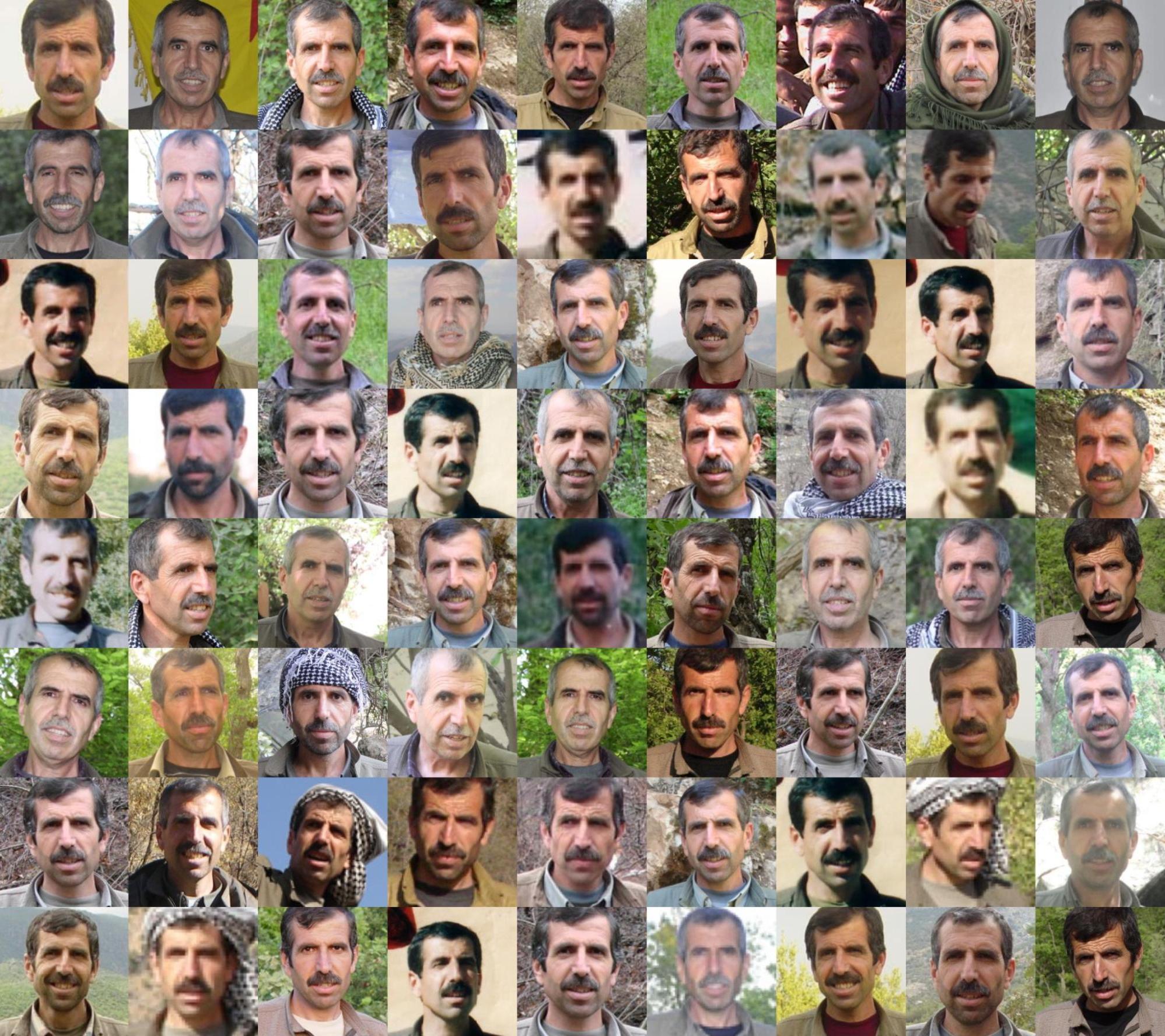}
{\scriptsize The clustering algorithm identifies Erdal in 72 unconstrained images. He is identified across a number of different lighting conditions and age ranges. \par}
\end{minipage}
  \label{figure6}
\end{figure}

The quantitative accuracy statistics provided above are enabled by the fact that images are sorted by obituary. As a final step, a mosaic image composed of all of the face tiles in a given cluster is created, enabling a visual assessment of clustering performance where image data are completely unstructured. Figure \ref{figure6} shows an example of the face tile mosaic generated for Bahoz Erdal, the leader of the PKK’s armed wing. The clustering process manages to identify him across an impressive range of lighting conditions, pose variations, image resolutions, ages, and hair styles without any false positives. 

\subsection{Generating a Network}

Having identified individuals across multiple images, a network $G=(N,E)$ is generated wherein each node N is an individual, and each edge E connecting nodes indicates that the individuals appeared in a photograph together. The PKK Image Co-Appearance Network (PICAN) is weighted, meaning that an edge linking two vertices will take on the value of the number of co-appearances between the two individuals. For example, if individuals a and b appear together in one photograph, edge $(a,b)=1$. If individuals c and d appear together in 10 photographs, edge $(c,d)=10$. The degree $k_{i}$ of node i reflects the number of edges linking that node to others. These edges are undirected. 

\begin{figure}[h]
  \caption{PKK Image Co-Appearance Network (PICAN)}
      \begin{minipage}{\columnwidth}
  \includegraphics[width=\columnwidth]{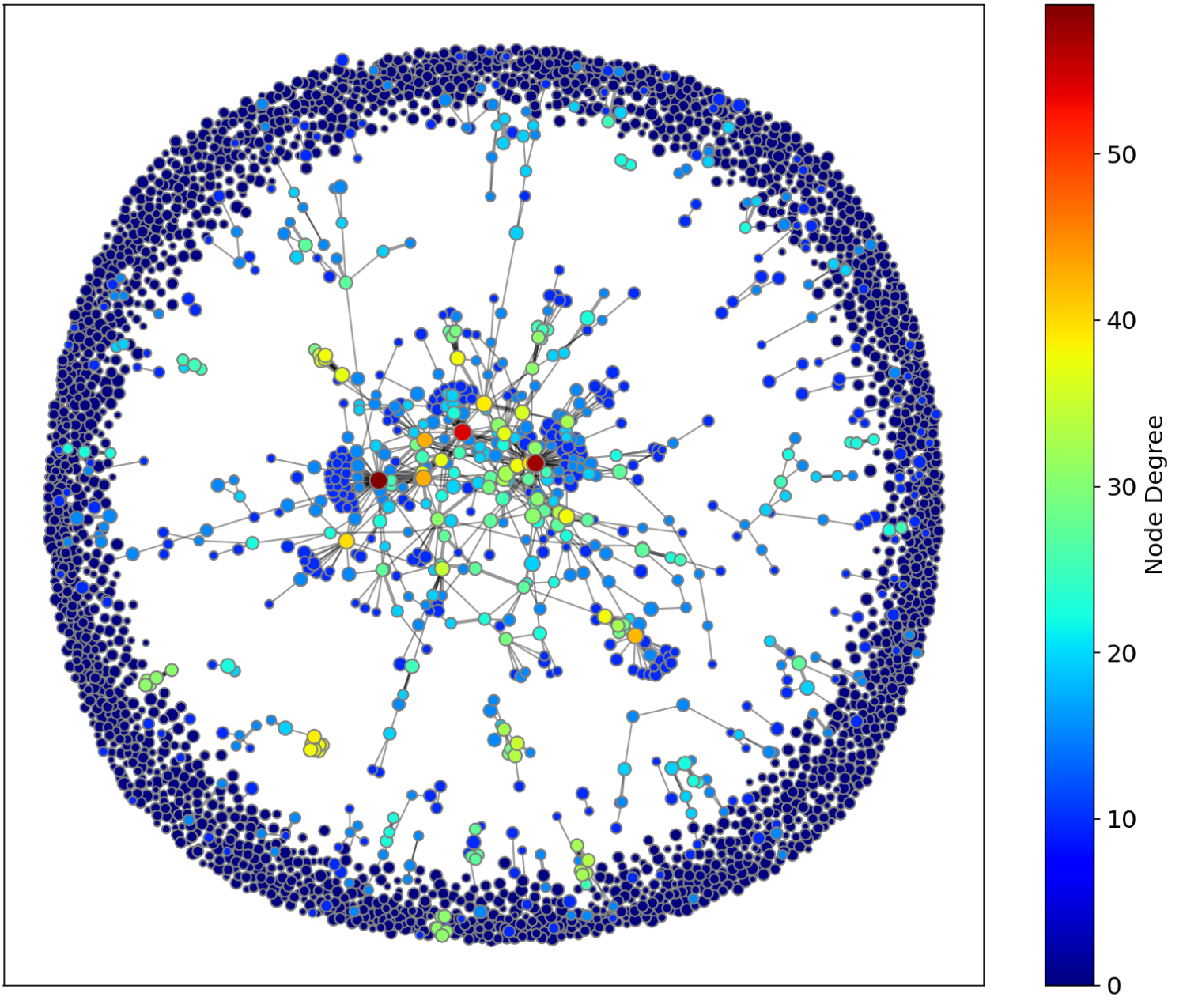}
{\scriptsize This figure displays a co-appearance network generated from nearly 20,000 images published online by the PKK. Each node is a cluster of faces corresponding to an individual (see Figure \ref{figure5}). Edges link nodes if the individuals appear together in a photograph, and are weighted by the number of co-appearances. Nodes are sized based on the number of pictures, and colored according to the number of co-appearances. 
\par}
\end{minipage}
  \label{figure7}
\end{figure}

In Figure \ref{figure7}, a node’s size is dictated by the number of images of an individual, while the node’s color reflects the number of co-appearances. The network has 2999 nodes and 991 edges. A large number of these nodes are isolates, meaning that they have no edges (k=0), making up the ring of blue nodes around the graph. These isolates derive from obituaries in which there are either a small number of images or which only contain portraits. The largest connected component (LCC) comprises 491 nodes, and is represented as the densely connected object in the center of the graph. Though there are also a number of smaller connected components, subsequent analysis will focus primarily on the LCC. Having generated the co-appearance network, the rest of this paper analyzes the properties of the PICAN in reference to external information about how the PKK is organized. 

\section{Functional and Factional Divisions}
\label{sec:qual}

\begin{figure*}[!h]
  \caption{PKK Leaders: the 15 Most Central Nodes}
      \begin{minipage}{\textwidth}
  \includegraphics[width=\textwidth]{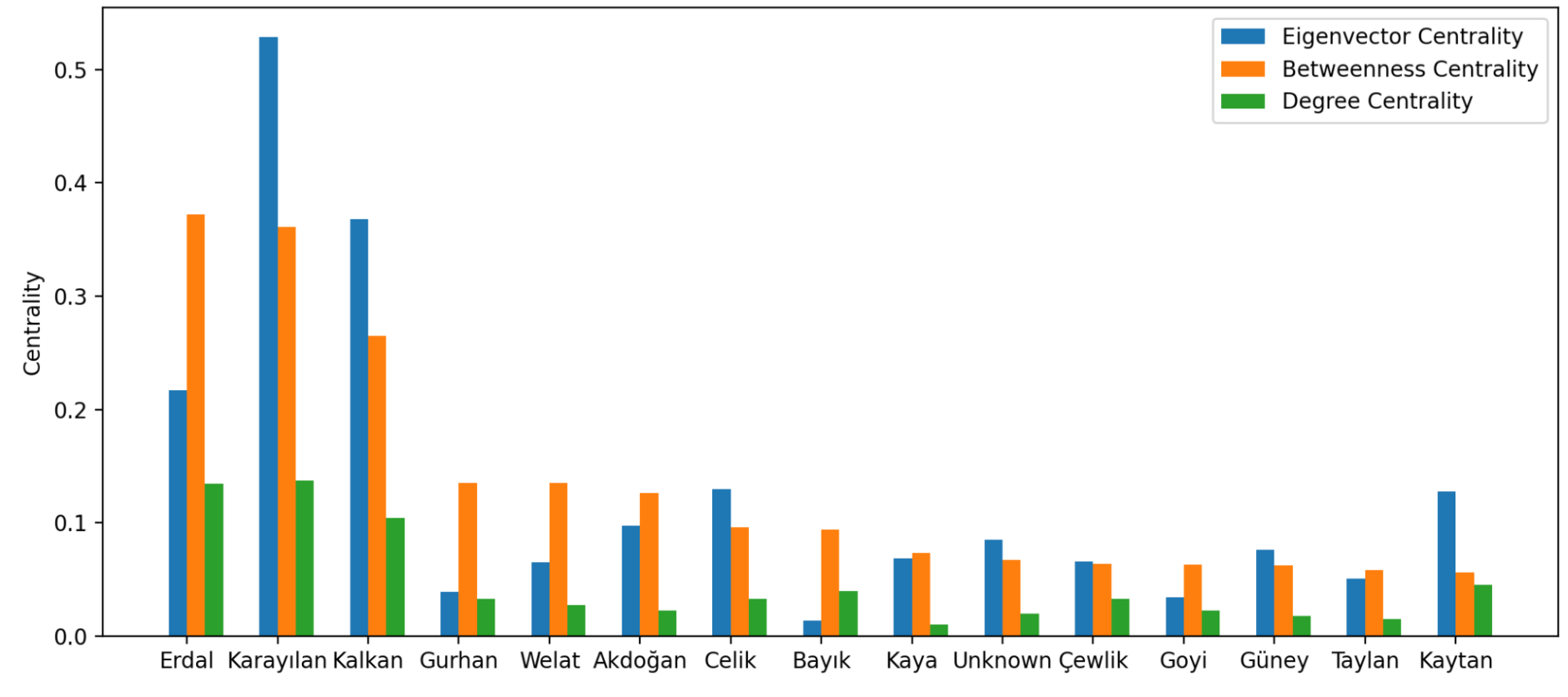}
{\scriptsize The 15 most central nodes in the PICAN correspond to known leaders of the PKK. Different measures of centrality shed light on the nature of their image co-appearances, which correspond to functional and factional divisions within the PKK. 
\par}
\end{minipage}  \label{figure8}
\end{figure*}

The performative nature of the photographs enables a rich qualitative interpretation of network statistics. Degree centrality simply measures the number of neighbors that a node has, which in this case is simply the number of times they appeared in photographs with others. Eigenvector centrality takes into account how well connected a node’s neighbours are. A node with relatively low degree can have high eigenvector centrality if they are connected to well-connected nodes\citep{golbeck_introduction_2015}. As such, eigenvector centrality is often interpreted as a measure of popularity in social networks \citep{borgatti_analyzing_2018}. Betweenness centrality measures the fraction of shortest paths that pass through a given node \citep{jia_edge_2012}. Nodes with high betweenness centrality occupy a powerful structural position in the network as they function as a “broker or a bridge” between subgroups and the rest of the network \citep{hansen_analyzing_2011}. 

There are multiple mechanisms through which node degree in an image co-appearance network could act as a proxy for seniority. The most parsimonious explanation is that most high-ranking members of the PKK have simply been active in the field for longer. The longer an individual is active, the more opportunities they have to appear in images with others. Another likely reason linking node degree and rank relates to the occasions in which pictures are taken. As previously discussed, many photographs are taken at ceremonies in which high-ranking members are present such as training academy graduations. New recruits often pose with commanders during these ceremonies, thereby generating a large number of edges in the PICAN.

Figure \ref{figure8} displays the 15 most central nodes in the PICAN using all three measures, ordered by betweenness centrality. There are three nodes that stand out regardless of which measure of centrality is employed: Bahoz Erdal, Murat Karayilan, and Duran Kalkan. The node with the highest betweenness centrality in the PICAN is that of Bahoz Erdal, who has commanded the PKK’s armed wing, the HPG (Hêzên Parastina Gel) since its founding in 2004. Despite his seniority in the PKK’s military apparatus, he is not a member of the PKK’s Executive Committee. Images of Erdal date back decades, and he appears to be primarily posing with rank and file militants; he is frequently the only connection for many of his neighbouring nodes. As the PKK’s top military commander, his ubiquity in pictures taken in the field is to be expected. Though Karayilan has a slightly lower betweenness centrality than Erdal, his eigenvector centrality is more than twice as great. Karayilan is the current leader of the PKK, and frequently appears alongside other members of the PKK’s executive council and top commanders as well as rank-and-file members. Unlike Erdal, many of his neighbours are also high-degree nodes rather than rank-and-file members, inflating his eigenvector centrality. The same is true of Duran Kalkan, who is also a member of the Executive Committee. Intuitively, those who appear most frequently with others tend to be high-ranking individuals. However, a closer examination of those that do not have many co-appearances despite their high rank demonstrates that internal political discord is organically reflected in these photographs.

A particularly interesting node in this regard is that of Cemil Bayik, one of the founding members of the PKK and one of the three current members of the PKK’s Executive Committee alongside Karayilan and Kalkan \citep{stansfield_kurdish_2016}. He was considered the PKK’s “longtime number two” after founder Abdullah Öcalan, and was the top commander of the PKK’s former armed wing, the ARGK (Artêşa Rizgariya Gêle Kurdistan) \citep{gunter_continuing_2000}. Despite being a central figure in the PKK’s development, Bayik’s node in the PKK network is far less central than those of his peers in the Executive Committee, Karayilan and Kalkan. His eigenvector centrality is particularly low-- the lowest in the sample above, which is surprising given that the other Executive Committee members have the highest eigenvector centrality in the network by virtue of frequently posing with other high ranking individuals. Indeed, in this sample of nearly 20,000 photographs spanning decades, Bayik-- one of the highest ranking members of the PKK on paper-- does not appear in a single photograph with any of the other top leaders of the organization, including the two other members of the Executive Committee.

\begin{figure}[h]
  \caption{PKK Leadership in the PICAN LCC}
      \begin{minipage}{\columnwidth}
  \includegraphics[width=\columnwidth]{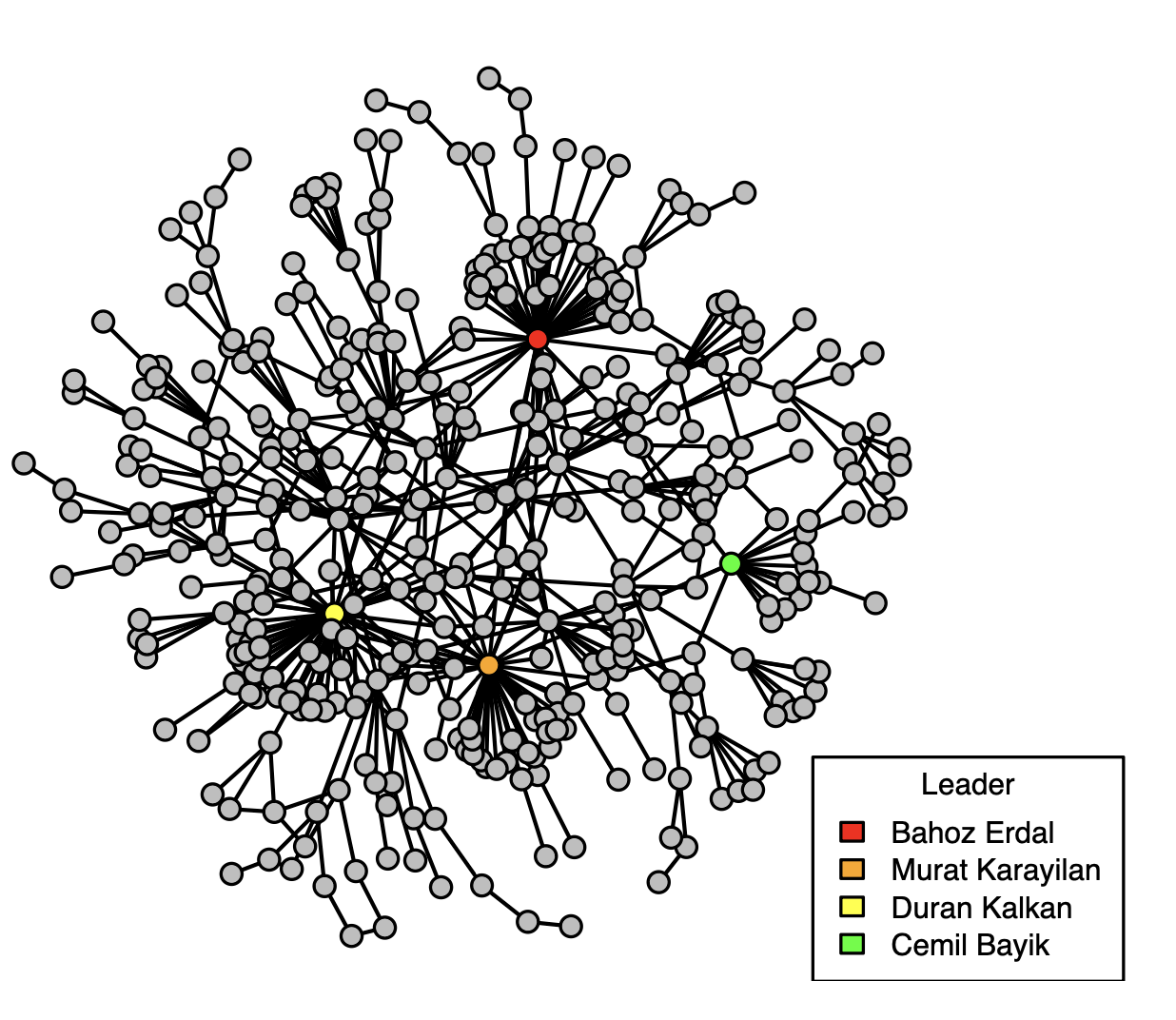}
{\scriptsize The position of four senior members of the PKK are shown in the PICAN. Karayilan, Kalkan, and Bayik are the three members of the PKK's Executive Committee. Erdal is the leader of the PKK's armed wing, the HPG. 
\par}
\end{minipage}    
\label{figure9}
\end{figure}

This discrepancy appears to be the result of a failed leadership struggle that occurred during the PKK’s reorganization following its military defeat in 1999. The PKK dissolved the military wing led by Bayik (the ARGK), and replaced it with a new wing called the HPG, led by Bahoz Erdal. Indeed, “when Bahoz Erdal became the head of the PKK’s military wing in June 2004, Bayik apparently found himself in a lesser position.” \citep[p. 65]{gunter_out_2014}. The reasons for Bayik’s ouster were likely related to his fraught relationship with the PKK’s founder; “Öcalan has criticized Bayik’s military leadership and ability to direct a guerrilla movement. According to Öcalan, Bayik prefers to stay behind the frontlines and has become involved in several controversial cases. For instance in 1992, fearing capture by the Turkish security forces, Bayik killed 17 wounded PKK militants in a cave.” \citep{uslu_leading_2008}. When the PKK resumed military operations in 2004, Bayik was consigned to diplomatic duties and spent much of his time in Iran, far from the battlefield. As a result, Bayik simply appears in far fewer photographs than his peers in the Executive Council. 

People who frequently appear with others in these photographs tend to be high ranking. Indeed, the strength of the relationship between how socially embedded an individual is (i.e. their centrality in the network) and their influence within the PKK is evidenced by the fact that the PICAN even captures the effect of internal political struggles; the discrepancy between Cemil Bayik’s status as a member of the PKK’s Executive Committee and his relatively low centrality in the PICAN reflects his marginalization following a failed leadership struggle. Rather than being a source of bias, the performative nature of photography in the context of militant photographs is itself a rich source of contextual information that can be interpreted through an examination of the nature of the co-appearances.

\section{Node Centrality and Rank}
\label{sec:rank}

\begin{table*}[h!]
\centering
\begin{threeparttable}
  \caption{Summary Statistics: Matching PICAN Nodes with Government Watchlists}

\begin{tabular}{lllllll}

\toprule
\hline\hline

 &                &     Red &    Blue &   Green &  Orange &    Grey \\
 \hline

\midrule
Panel A:            &     Reward (TL, 000) &   10000 &    3000 &    2000 &    1000 &     500 \\
Matching Statistics &  Listed Individuals &     178 &      49 &      73 &     158 &     581 \\
                    &             Matched &      28 &       9 &       6 &      20 &      59 \\
                    &           \% Matched &    15.7 &    18.4 &     8.2 &    12.7 &    10.2 \\
\hline
Panel B:            &         Image Count &   14.17 &   12.17 &     3.0 &    7.67 &    6.95 \\
Network Centrality  &          \% Isolates &  13.79 &  58.33 &  71.43 &  85.71 &  84.48 \\
                    &              Degree &  0.0235 &  0.0068 &  0.0014 &  0.0014 &   0.001 \\
                    &         Eigenvector &  0.0629 &  0.0332 &  0.0102 &  0.0008 &  0.0071 \\
                    &         Betweenness &  0.0571 &  0.0222 &     0.0 &  0.0017 &  0.0015 \\
\bottomrule
\hline
\hline

\end{tabular}
\begin{tablenotes}
\scriptsize
\item

The columns in this table indicate the colors of five wanted persons lists maintained by the Turkish Government. Panel A reports the value of the reward offered for individuals on each list, the number of listed individuals, and the number and proportion of PICAN nodes that were matched with these individuals. Panel B reports  attributes for matched nodes in each list, including the average number of images they appear in, the proportion that are isolates, as well as three measures of centrality. 
\end{tablenotes}
\label{table1}
\end{threeparttable}
\end{table*}
The previous section established that image co-appearance can highlight the top leadership of a rebel group, and even shed light on political divisions therein. However, the link between photogenicity and rank may only exist for the highest ranking (and thus, most salient) members of an insurgent group: while new recruits may clamor for a photograph with the supreme leader, a platoon commander may inspire less enthusiasm. This section matches 126 individuals from wanted lists maintained by the Turkish government with nodes in the PICAN to test whether there is a relationship between an individual’s rank within the PKK (as measured by the reward offered for their capture) and their corresponding node’s position in the PICAN. Linear regression models suggest that over half of the variation in the monetary reward offered for an individual can be explained by image co-appearances alone. Exponential Random Graph Models show that higher ranking individuals are significantly more likely to form ties in the PICAN. These results are not simply driven by the top leadership posing with new recruits; In both sets of models, the relationship between seniority and centrality is robust to the exclusion of the 28 highest-ranking officials (those on the Red List), suggesting that photogenicity gradually increases with rank.

The Turkish government maintains lists of wanted persons (including those killed, captured, and at large), offering monetary rewards for information leading to their capture. Individuals are assigned into one of five color coded wanted lists (red, blue, green, orange, and grey), with reward amounts ranging from 10 million Turkish Lira for individuals on the Red List to 500,000 TL for individuals on the Grey List. The value of the reward offered for an individual’s capture (i.e. the color of the wanted list that they are assigned to) is largely a function of the Turkish government’s assessment of that individual’s rank within the PKK. Article 6 (2) of the “Regulation on Rewards to be Given to Those Who Help in Discovering Terrorist Crimes” states that individuals are assigned to one of these five lists by “[...] grouping them according to their hierarchical position in the terrorist organization and/or the weight of the consequences of their actions, and by specifying the maximum amount of reward that can be given to those in each group” (Turkish Ministry of Interior, 2022). This legislation, available on the wanted website, further specifies that the highest reward amounts are offered for individuals on the Red List, who are deemed to be “senior manager[s] in a terrorist organization”. Though no precise information is given regarding how exactly individuals are assigned to different lists, individuals on the Blue List seem to represent the second echelon judging by the relatively small number of individuals in this category (49), as well as the reward amount (3 million). The Green and Orange Lists likely correspond to PKK members who are believed to have some level of managerial responsibility that justifies offering rewards of 2 million and 1 million, respectively. The Grey List offers the lowest reward amount and is by far the largest group, with over 500 listed individuals, which may suggest that they are known to be members of the PKK, but do not occupy positions of power. 

\begin{figure*}[h]
  \caption{PICAN Nodes Matched with Wanted Lists}
      \begin{minipage}{\textwidth}
  \includegraphics[width=\textwidth]{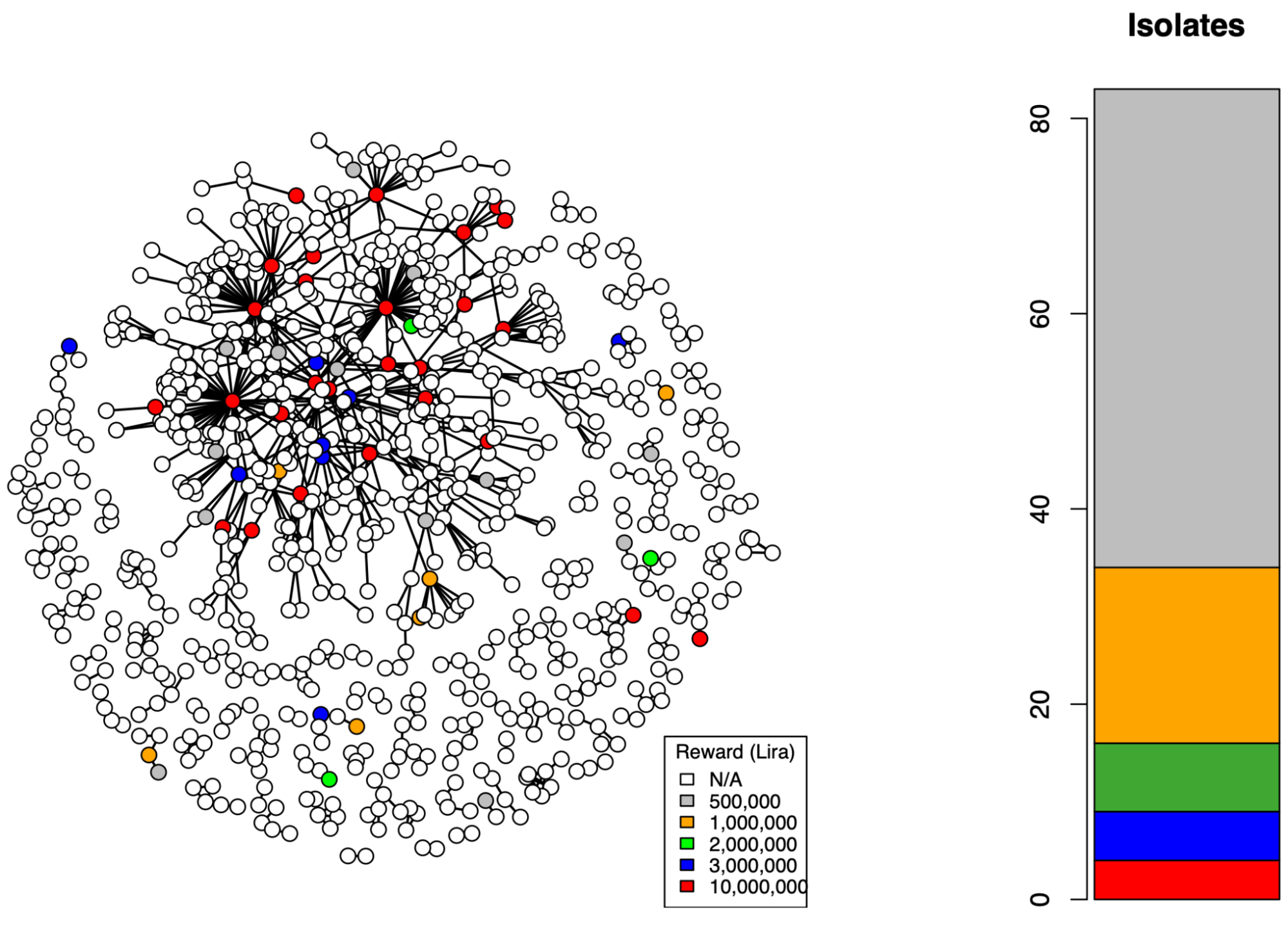}
{\scriptsize In the graph of connected components on the left, nodes in the PICAN are matched with wanted persons and colored according to the value of the reward offered for their capture. The bar on the right counts the number of isolates that were matched at each wanted level. Individuals on the Grey and Orange Lists-- the least wanted categories-- tend to be isolates. Individuals on the higher-priority Red, Blue, and Green lists tend to be well connected. 
\par}
\end{minipage}  
  \label{figure10}
\end{figure*}

To examine the relationship between network centrality measures and rank, individuals from the wanted lists were matched with nodes in the network using facial recognition and text-based matching. Each entry on the wanted lists contains a photograph of the individual, their full name, date and place of birth, and the organization they are alleged to be a member of. Facial recognition was performed using the individual’s photo on the wanted website, employing the same euclidean distance threshold (0.39) as the one used for in the clustering step of the construction of the PICAN network. Text-based matching was also conducted by identifying exact matches in terms of first name and last name between both datasets. Both facial and text-based matches were verified manually, yielding a total of 126 confirmed matches. 

Figure \ref{figure10} shows the position of nodes that were matched with records on the wanted list, coloring them according to their wanted level. The graph of connected components on the left shows that individuals on the Red and Blue lists occupy highly central positions in the network. The bar chart on the right suggests that very few of these individuals are isolates. Conversely, individuals on the Grey and Orange lists are largely absent from the graph of connected components or occupy relatively peripheral positions. The bulk of these individuals have no co-appearances at all. 

The number and relative proportions of individuals identified in each wanted category are reported in Panel A of Table \ref{table1}. The generally low match rates can be attributed to the inherent challenges of identifying members of a clandestine organization using facial recognition and imagery from two different sources. In particular, poor image quality posed a significant challenge for facial recognition-- most of the images on the wanted list website were unusable due to being stretched to fit the dimensions of the image element, and only one image was available for each individual. Failed matching due to image quality generates data that is Missing Completely At Random (MCAR), which simply reduces the sample size. However, differing match rates between wanted lists raises concerns about selection bias. The highest match rates were among those in the most wanted categories-- 16\% of the individuals on the Red List and 18\% of those on the Blue List were matched to nodes in the PICAN. Lower match rates were observed for the Green, Orange, and Grey lists, which could indicate that lower-ranking individuals are underrepresented in the obituary images. However, the lowest-ranking group (the Grey List) has the largest total number of matches, over twice as many as the group with the next-highest number of matches.

Panel B of Table \ref{table1} reports the average values of several network statistics for each group. On average, individuals on the Red List appear in 14 photos, while individuals on the Grey List appear in just 7. The proportion of nodes in each group that are isolates decreases as the reward amount increases; only 14\% of nodes in the Red List were isolates, increasing to 84\% in the Grey List. The same is true of all three measures of centrality. The summary statistics and the graph in Figure \ref{figure10} seem to suggest a positive relationship between node centrality and an individual’s rank within the PKK as determined by the value of the reward placed on them by the Turkish government. 

\subsection{Linear Models}

To further explore the relationship between node centrality and rank, I specify a simple OLS regression model of the following form:

$${Y}_{i}=\beta_0+\beta_1{x}_{i}+\epsilon_{i}$$

Where y is the reward amount (in thousands of Turkish Lira) offered for individual i, and x is a measure of node centrality. Because measures of centrality are highly correlated with each other, a separate regression is run for each metric. Three common measures of centrality are used: degree, eigenvector, and betweenness. These are standardized by the number of images an individual appears in to emphasize social embeddedness rather than photogenicity without introducing multicollinearity between an image count variable and the centrality measures. For simple degree centrality, this effectively yields a measure of co-appearances per image. Table \ref{table2} summarizes the results of these regressions.  

\begin{table}[h!]
\begin{threeparttable}
\caption{OLS Regression Coefficients for Standardized Network Centrality Measures and Reward Value}
\begin{center}
\begin{tabular}{p{1.8cm}p{1.2cm}p{1.2cm}p{1.2cm}p{1.2cm}}
\hline
 & Model 1 & Model 2 & Model 3 & Model 4 \\
\hline
(Intercept) & $2051.32^{***}$ & $1466.35^{***}$    & $2442.66^{***}$   & $1950.80^{***}$    \\
            & $(445.61)$      & $(269.17)$         & $(339.06)$        & $(297.39)$         \\
Image Count & $113.63^{***}$  &                    &                   &                    \\
            & $(33.00)$       &                    &                   &                    \\
Degree      &                 & $3134293.53^{***}$ &                   &                    \\
            &                 & $(258269.53)$      &                   &                    \\
Eigenvector &                 &                    & $352246.85^{***}$ &                    \\
            &                 &                    & $(69378.62)$      &                    \\
Betweenness &                 &                    &                   & $1117667.87^{***}$ \\
            &                 &                    &                   & $(124487.94)$      \\
\hline
R$^2$       & $0.09$          & $0.54$             & $0.17$            & $0.39$             \\
Adj. R$^2$  & $0.08$          & $0.54$             & $0.17$            & $0.39$             \\
Num. obs.   & $126$           & $126$              & $126$             & $126$              \\
\hline
\multicolumn{5}{l}{\scriptsize{$^{***}p<0.001$; $^{**}p<0.01$; $^{*}p<0.05$}}
\end{tabular}
\begin{tablenotes}
\scriptsize
\item
The dependent variable in all models is the value of the reward placed on an individual by the Turkish Government. The independent variable in Model 1 is the number of photographs an individual appears in. Models 2, 3, and 4 use different measures of node centrality divided by the number of images that individual appears in. 
\end{tablenotes}
\label{table2}
\end{center}
\end{threeparttable}
\end{table}

\begin{figure}[!h]
  \caption{Network Centrality and Wanted Level}
      \begin{minipage}{\columnwidth}
  \includegraphics[width=\columnwidth]{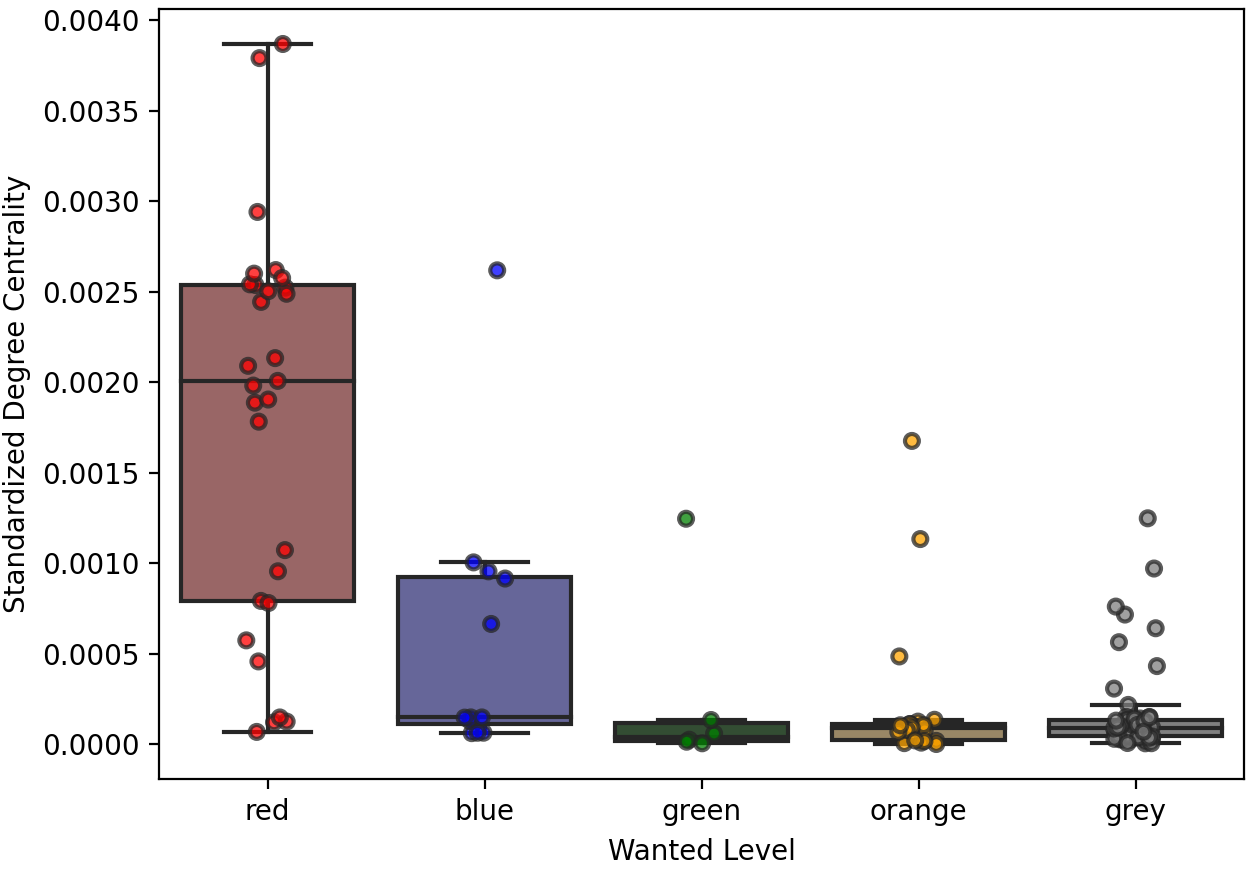}
{\scriptsize The colors represent wanted lists, which offer monetary rewards for an individual's capture. They are arranged in descending order of reward value: 10m, 3m, 2m, 1m, and 0.5m. The Y axis represents degree centrality divided by the number of images an individual appears in. Individuals on the red and blue lists tend to be far more central than those on the lower-priority lists. 
\par}
\end{minipage}    
\label{figure11}
\end{figure}

Model 1 regresses the count of images an individual is identified in with the value of the reward placed on them. Though not a measure of centrality per se, this provides useful contextual information on whether higher ranking individuals appear in more photographs than those of lower rank. Indeed, the results from the first column indicate that for each additional photograph an individual appears in, the reward amount increases by 114,000 Turkish Lira on average. However, less than 10\% of the variation in reward amounts is explained by the number of photos an individual appears in. Models 2, 3, and 4 show strong positive associations between degree, eigenvector, and betweenness centrality and reward values. The model with the greatest explanatory power relies on simple degree centrality, achieving an $R^2$ value of 0.54; in other words, over half of the variation in the bounty price placed on an individual can be explained by the frequency with which they appear with others in photographs. The boxplot in Figure \ref{figure11} shows the relationship between standardized degree centrality and wanted level.

To ensure that these results are not entirely driven by extreme values, the models in Table \ref{table2} are re-run excluding the 28 individuals who appear on the Red List. The top leaders of rebel groups are often well known, but information on the structure of rebel groups below the level of top leadership is much harder to come by; second and third order commanders are less prolific. By excluding the top leadership from the regressions, the results in Table \ref{table3} assess whether node centrality measures not only distinguish the top echelon from the rank-and-file, but whether they can distinguish lower ranks as well. 

\begin{table}[h!]
\begin{threeparttable}
\caption{OLS Regression Coefficients for Network Centrality Measures and Reward Value, Excluding Red List}
\begin{center}
\begin{tabular}{p{1.8cm}p{1.2cm}p{1.2cm}p{1.2cm}p{1.2cm}}
\hline
 & Model 1 & Model 2 & Model 3 & Model 4 \\
\hline
(Intercept) & $672.40^{***}$ & $913.71^{***}$  & $932.82^{***}$ & $901.13^{***}$   \\
            & $(153.90)$     & $(86.44)$       & $(87.19)$      & $(80.83)$        \\
Image Count & $44.96^{*}$    &                 &                &                  \\
            & $(17.22)$      &                 &                &                  \\
Degree      &                & $53811.91^{**}$ &                &                  \\
            &                & $(16026.47)$    &                &                  \\
Eigenvector &                &                 & $8802.63^{**}$ &                  \\
            &                &                 & $(3108.03)$    &                  \\
Betweenness &                &                 &                & $27185.59^{***}$ \\
            &                &                 &                & $(5722.21)$      \\
\hline
R$^2$       & $0.07$         & $0.11$          & $0.08$         & $0.19$           \\
Adj. R$^2$  & $0.06$         & $0.10$          & $0.07$         & $0.18$           \\
Num. obs.   & $97$           & $97$            & $97$           & $97$             \\
\hline
\multicolumn{5}{l}{\scriptsize{$^{***}p<0.001$; $^{**}p<0.01$; $^{*}p<0.05$ }}
\end{tabular}
\begin{tablenotes}
\scriptsize
\item
The models in this table are identical to those presented in Table \ref{table2}, except that the 28 individuals in the most wanted category (the Red List) are excluded. Results are robust to this exclusion. 
\end{tablenotes}
\label{table3}
\end{center}
\end{threeparttable}
\end{table}

Model 1 shows a substantial decrease in the effect size and significance of the relationship between the number of photos an individual appears in and their reward value when the red list is excluded. This reflects some of the idiosyncrasies of picture-taking in the PKK: foot soldiers can have a relatively large number of portrait photos, but higher ranking individuals (particularly those that are still alive) rarely appear alone. Despite reductions in effect size, significance, and model fit across the board, Models 2-4 nonetheless display a consistent relationship between node centrality and wanted level even when the highest echelon is removed from the analysis. 

\subsection{Exponential Random Graph Models}

Though \cite{borgatti_analyzing_2018} note that hypothesis testing using node-level attributes is mostly conducted via standard regression models in the literature, the use of network data violates the assumption that observations are independent. This problem is overcome by Exponential-family Random Graph Models (ERGMs), which enable an assessment of the statistical likelihood of observing certain network configurations, including the influence of nodal attributes on tie formation. ERGMs treat the observed network (in this case, the PICAN) as one realization from a set of possible networks with similar core characteristics \citep{robins_introduction_2007}. The general form specifies the probability of the observed network as a function of a set of network features that may occur more or less frequently than expected by chance: 

$$P(Y=y)=\frac{exp(\theta' g(y))}{k(\theta)} $$

Where $Y$ is the random variable for the state of the network, and $y$ is the realization thereof. The right hand side specifies a vector of coefficients $\theta$ for model statistics $g$ of the network $y$, normalized by the sum of all possible networks $k(\theta)$ \citep{statnet_introduction_2022}. These networks are generated via Markov Chain Monte Carlo maximum likelihood estimation (MCMCMLE). If more or less of a certain network configuration is present in the observed network than in the simulated one, the parameter corresponding to that configuration is iteratively adjusted in subsequent simulations until the simulated network closely approximates the observed one \citep{ellwardt_who_2012}. For example, given a parameter for the number of edges involving female nodes, a positive value would indicate that women have a greater propensity to form ties in the PICAN than in a random graph. 

Table \ref{table4} reports the results from four ERG Models which assess whether high-ranking PKK members (i.e., those fetching higher bounties) are more likely to co-appear in images than lower ranking individuals. Models 1 and 2 utilize the full PICAN and include a parameter for isolates, while models 3 and 4 exclude isolates altogether. The inclusion of a triadic close term in the full PICAN leads to model degeneracy, so it is only included in the analysis of connected components. 

\begin{figure*}
  \caption{ERGM Goodness of Fit Diagnostics}
      \begin{minipage}{\textwidth}
  \includegraphics[width=\textwidth]{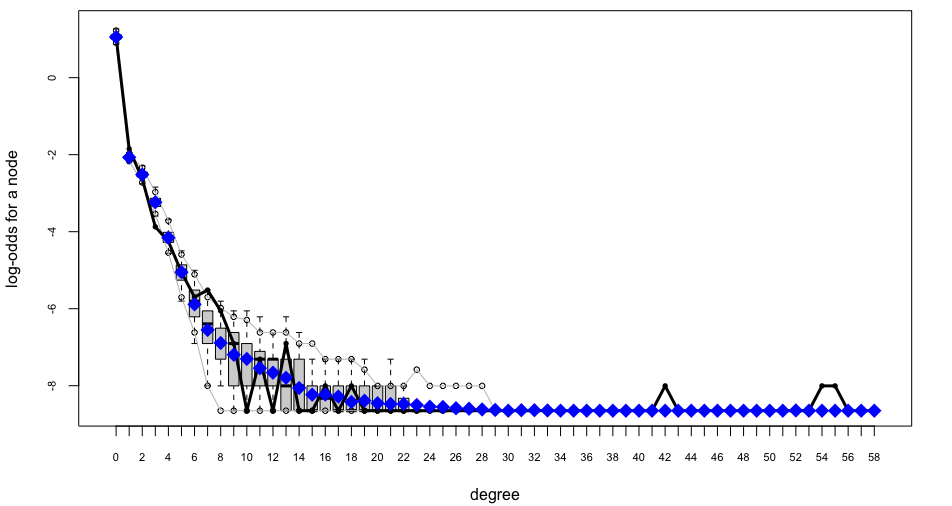}
{\footnotesize The empirical distribution is represented as a solid black line, while the simulated distributions are indicated by the boxplots. Though three very high degree nodes-- the PKK's main leaders-- consistently defy simulation, the rest of the observed degree distribution is generally within the bounds of the simulated models. 
\par}
\end{minipage}     
\label{GOF}
\end{figure*}

All models converged successfully, indicating that the final simulated network in each case was sufficiently close to the observed network in terms of the included parameters. To test the extent to which the Exponential Random Graph Model presented in Table \ref{table4} fits the data, it is convention to observe how well the model reproduces network properties that are not included as parameters in the estimation. The main hypothesis being tested in the model is monadic in nature (whether the nodal attribute of reward value influences the likelihood of tie formation). As such, similar to \cite{stys_trust_2022}, Figure \ref{GOF} compares a node-level network statistic-- the degree distribution-- between the observed and simulated networks. The simulated networks largely reproduce network statistics in the PICAN that were not included as terms in the ERGMs, suggesting that the models faithfully approximate the observed network. 

\begin{table}
\begin{threeparttable}
\caption{Effect of Reward Value on the Likelihood of Edge Formation: Parameter Estimates from ERGMs}
\begin{center}
\begin{tabular}{p{3cm}p{1cm}p{1cm}p{1cm}p{1cm}}
\hline
 & Model 1 & Model 2 & Model 3 & Model 4 \\
\hline
Edges              & $-9.69^{***}$ & $-9.53^{***}$ & $-7.90^{***}$ & $-7.55^{***}$ \\
                   & $(0.20)$      & $(0.33)$      & $(0.16)$      & $(0.33)$      \\
Isolates           & $2.09^{***}$  & $2.08^{***}$  &               &               \\
                   & $(0.08)$      & $(0.08)$      &               &               \\
Triadic Closure    &               &               & $1.72^{***}$  & $1.71^{***}$  \\
                   &               &               & $(0.06)$      & $(0.06)$      \\
Age                & $0.04^{***}$  & $0.04^{***}$  & $0.02^{***}$  & $0.02^{***}$  \\
                   & $(0.00)$      & $(0.00)$      & $(0.00)$      & $(0.00)$      \\
Gender             & $0.13^{**}$   & $0.12^{**}$   & $-0.00$       & $-0.01$       \\
                   & $(0.05)$      & $(0.05)$      & $(0.05)$      & $(0.05)$      \\
Reward             & $0.14^{***}$  &               & $0.07^{***}$  &               \\
                   & $(0.01)$      &               & $(0.01)$      &               \\
Wanted (Unmatched) &               & $-0.08$       &               & $-0.19$       \\
                   &               & $(0.14)$      &               & $(0.15)$      \\
Wanted (1M)        &               & $0.04$        &               & $0.09$        \\
                   &               & $(0.23)$      &               & $(0.24)$      \\
Wanted (2M)        &               & $-0.27$       &               & $-0.73$       \\
                   &               & $(0.48)$      &               & $(0.52)$      \\
Wanted (3M)        &               & $0.59^{**}$   &               & $0.24$        \\
                   &               & $(0.19)$      &               & $(0.19)$      \\
Wanted (10M)       &               & $1.30^{***}$  &               & $0.51^{**}$   \\
                   &               & $(0.16)$      &               & $(0.16)$      \\
\hline
AIC                & $14129.42$    & $14133.88$    & $10476.49$    & $10471.77$    \\
BIC                & $14196.00$    & $14253.72$    & $10529.52$    & $10567.23$    \\
Log Likelihood     & $-7059.71$    & $-7057.94$    & $-5233.24$    & $-5226.89$    \\
\hline
\multicolumn{5}{l}{\scriptsize{$^{***}p<0.001$; $^{**}p<0.01$; $^{*}p<0.05$ }}
\end{tabular}
\begin{tablenotes}
\scriptsize
\item
Models 1 and 2 use the full PICAN and include a paramter for isolates. Models 3 and 4 exclude isolates, but include a parameter for triadic closure (Geometric Edgewise Shared Partners or GWESP). Age and Gender are node-level attributes. "Reward" is a continuous node-level attribute indicating the value of reward placed on an individual by the Turkish government. "Wanted" is a categorical version of this variable, with the reward for each category indicated in parentheses. 
\end{tablenotes}

\label{table4}
\end{center}
\end{threeparttable}
\end{table}

All models also include apparent age and gender as control variables. Gender has a modest but significant effect in the full network: women are 1.14 times more likely to form ties than men, which is consistent with the prior observation that women appear in more images\footnote[2]{$e^{0.13}=1.14$}. This likely reflects the performative nature of many of these photographs, which seek to emphasize the role of women in the PKK. Age also increases the likelihood of edge formation, which may result from the fact that older members have simply had more time to appear in pictures. Alternatively, age could also be proxying for seniority. Age and rank are likely to be correlated in any military setting: infantrymen are rarely old, generals are rarely young, and the PKK’s known leaders are all above the age of 60. Despite this, the “Reward” variable remains positive and significant in both models, though the magnitude of the effect is diminished by the addition of controls. 

The main variable of interest in all models is the value of the reward placed on an individual by the Turkish government. Models 1 and 3 include a continuous version of this variable, denoting the value of the reward offered for an individual in millions of Turkish Lira. In both the full network and the restricted sample, nodes corresponding to individuals with a higher reward value have a significantly higher probability of forming edges. In the full network, a node on the Red List is 10.92 times more likely to form a tie than a node on Grey List, ceteris paribus\footnote[3]{$(10 \times e^{0.14})-(0.5 \times e^{0.14})= 10.92$}. This effect is halved when isolates are excluded, likely due to the fact that 85\% of the nodes on the Grey list are isolates. 

Models 2 and 4 treat the reward variable as categorical rather than continuous. This enables the estimation of a separate effect for each reward amount and allows us to set an appropriate base category; though there are five different wanted lists corresponding to different reward amounts, this variable contains a sixth category: PICAN nodes that were not matched to wanted individuals. This constitutes an inappropriate reference group because information about these individuals’ rank is unknown; as such, each level of the “Reward” variable is compared against individuals on the Grey list (0.5M TL). Individuals on the Red and Blue wanted lists (fetching rewards of 10M TL and 3M TL, respectively) are significantly more likely to co-appear in images than individuals on the Grey List in the full network. However, the coefficient on the Blue List category becomes insignificant in Model 4, due to the exclusion of isolates (and thus over 85\% of the observations in the Grey List reference category). Despite the categorical nature of the variable, the ratio between the coefficients and the reward amounts of Red and Blue list members is remarkably close: the reward for an individual on the Red List is triple that offered for someone on the Blue List, and in both models the coefficient of the Red List is roughly double that of the Blue List. This aligns with the OLS results in Table \ref{table2}, suggesting that co-appearance not only identifies the top echelon of the PKK’s command structure, but lower ranking officers as well. However, individuals on the Green and Orange lists (1M and 2M) did not have a significantly higher likelihood of co-appearance relative to the Grey list, though this is likely due to the small number of observations in these categories.

The absence of a significant relationship between membership of the Green and Orange lists and likelihood of edge formation suggest that the relationship between social embeddedness and node centrality is not consistent across the full range of the data. Indeed, the lower match rates between individuals on these two lists and nodes in the PICAN imply that some lower-ranking officers may be disproportionately absent from the obituary photos altogether. Nevertheless, results from this section demonstrate a general relationship between an individual’s rank within the PKK (measured in terms of reward value) and the centrality of their corresponding node in the PICAN. Results from the linear regressions indicate that over half of the variation in reward values can be explained by the frequency with which an individual appears with others in photographs. All three of the centrality measures vastly outperform image count in terms of explaining variation in reward values, suggesting that social embeddedness is a better proxy for rank than photogenicity. ERGM results are consistent with these findings, with higher value individuals significantly more likely to form edges in the PICAN. Importantly, this relationship persists across all models even when the highest-ranking officials were excluded, meaning that results are not driven purely by the propensity of new recruits to pose with the group’s leader. 

An important caveat to this analysis is that the PICAN is only an approximation of the PKK’s structure. Co-appearances in these images can reflect a number of forms of social interaction, from friends sharing a meal to a new recruit posing with their commander. Edges do not provide any direct information on the nature of the relationship between two individuals, other than the fact that they crossed paths. At their most basic level, edges in the PICAN simply reflect the number of social contacts an individual has had. In aggregate, the quantity and nature of these interactions correlates strongly with an individual’s rank, allowing for the creation of a network that approximates the known structure of the PKK. It is important to note, however, that some individuals can be important to the organization without being highly socially embedded or visible in photographs from the field, and conversely that some highly photogenic individuals may be of low systemic importance. 

\section{Network Structure and Counterinsurgency Tactics}
\label{sec:robustness}
A significant gap between the theoretical and empirical literature on the application of SNA to the study of insurgent groups involves the relationship between network topology and robustness to counterinsurgency strategies \citep{helfstein_covert_2011}. This section seeks to assess the extent to which a co-appearance network generated from militant photographs can help to fill this gap.

Despite the strong association between rank and centrality, the PICAN does not represent an organogram of the PKK; it is a representation of how the PKK chooses to present itself to the world. This exaggerates certain characteristics, such as the group’s female cadres, while understating others such as lower-level commanders, the “dishonorably discharged”, and the politically marginalized. Nevertheless, given a large enough sample it seems plausible that an image co-appearance network would capture the general structural characteristics of an insurgent group-- the rough number and relative proportions of individuals at various ranks, and the relationships between them. In what follows, I analyse the topology of the PICAN in reference to historical developments in the conflict between the PKK and the Turkish government. 

\cite{unal_military_2016} identifies two main phases in Turkish military strategies towards the PKK: ground war and targeted killings. Until the year 2000, the Turkish military largely waged a conventional ground war against the PKK. Under this doctrine, the Turkish government conducted large scale military operations, armed roughly 60,000 civilians to act as paramilitary force and burned as many as 4,000 Kurdish villages to the ground \citep{gurcan_arming_2014, filkins_kurds_2003}. Though this strategy culminated in the PKK declaring military defeat in 2000, it failed to prevent the group’s resurgence only four years later. Following a period of low intensity conflict, large-scale hostilities resumed between the PKK and Turkish forces following the breakdown of a truce in 2015. This new phase in the conflict has been accompanied by a significant doctrinal shift on the part of the Turkish government towards a more targeted approach. Rather than relying on large military operations and poorly trained paramilitaries, the new approach relies heavily on the use of special forces, drone strikes, and intelligence gathering \citep{unal_military_2016}. Particularly since 2020, the Turkish government’s focus has been on targeting the leadership structure of the PKK \citep{uslu_leading_2008}. 

These various counterinsurgency strategies can be simulated, and their effects on the network can be assessed empirically. In general, the static robustness of a network to intentional attacks “analyses the ability of a system to maintain its connectivity after the disconnection or deletion of a series of targeted nodes. In this context, the connectivity of the resulting network is typically measured by the size of the largest connected component (LCC)” \citep[p. 1]{lordan_exact_2019}. This section proceeds with an empirical analysis of the two main counterinsurgency strategies in chronological order; first, the strategy of inflicting maximum casualties against the PKK through a ground war, and second, the doctrine of targeted strikes against high ranking members. 

\subsection{Ground War}

\begin{figure}
  \caption{PICAN Scale-Free Degree Distribution}
      \begin{minipage}{\columnwidth}
  \includegraphics[width=\columnwidth]{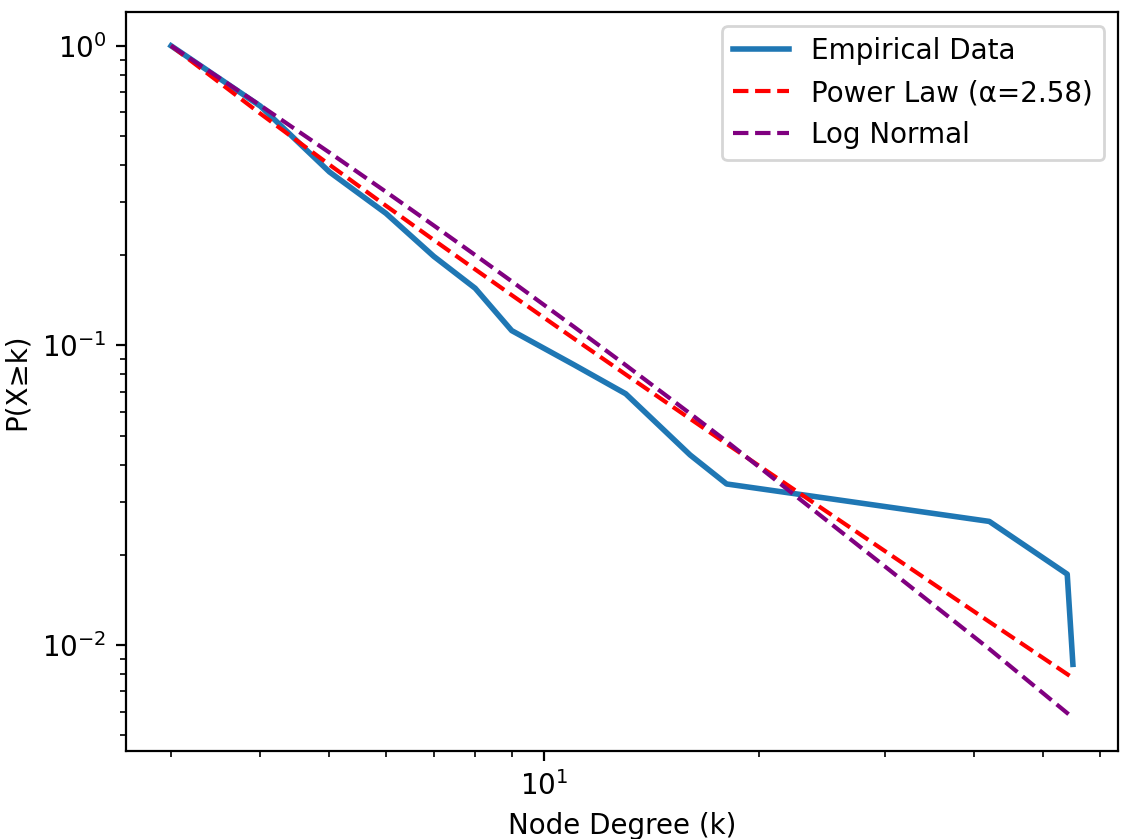}
{\scriptsize The PICAN's degree distribution (shown in blue) follows a power law of $k^{-2.58}$, making it a weakly scale-free network.
\par}
\end{minipage}    
\label{figure12}
\end{figure}

Much of the historical literature concerning first phase of the conflict between the Turkish Government and the PKK focuses on a puzzling dynamic: despite an aggressive military doctrine focused on engaging the PKK in a conventional war resulting in the group declaring military defeat, the PKK not only survived, but was able to regroup with remarkable speed. “By the early 2000s, the PKK was militarily weak, and the majority of its members were outside Turkey’s borders. However, after a few years of calm, on 1 June 2004, the PKK put an end to its unilateral ceasefire and once again began to attack civilian and military targets in Turkey. How was the PKK able to survive and rebuild itself in such a short period of time?” \citep[p. 727]{pusane_turkeys_2015}. Thus, in the early stages of the conflict, the Turkish government tried to defeat the PKK by inflicting the maximum number of casualties possible, with little concern for whom within the organization was being targeted. Translating this policy into the realm of social networks, this doctrine roughly equates to a strategy of random node removal. 

The robustness of a network to various types of node removal depends largely on its structure. A salient characteristic that can be observed in the PICAN is its highly uneven degree distribution. As shown by the previous section, despite the majority of nodes in the network being either isolates or low-degree, a small number of nodes have very high degree. A scale free network is a graph in which the nodes’ degree distribution follows a power law, where the probability $p(k)$ that a node has degree k is a function of $k^{- \alpha}$, such that $p(k)\propto k^{- \alpha}$, where $\alpha$ is a constant. Figure \ref{figure12} shows the empirical distribution of node degrees, compared with a power law distribution.

The empirical degree distribution appears to loosely follow a power law distribution of $k^{-2.58}$. According to \cite{broido_scale-free_2019}, “a network is deemed scale free if the fraction of nodes with degree k follows a power-law distribution $k^{- \alpha}$, where $\alpha>1$”, making the PICAN weakly scale-free. These types of networks have properties that are of direct relevance to counterinsurgency tactics; in particular, “a compelling characteristic of scale-free networks is that a significant fraction of nodes can be removed randomly without the network losing connectivity” \citep{verma_economics_2020}. This is because the vast majority of nodes have a low degree, while only a small number of nodes are responsible for the overall connectivity of the network. If nodes are removed at random, it is unlikely that a high-degree node would be removed. 

A counterinsurgency strategy based on inflicting the maximum amount of casualties against the PKK is analogous to a strategy of random node removal. This was the strategy pursued by the Turkish Armed Forces (TSK) in the 1990s, and it succeeded in weakening the PKK militarily to the point that “the PKK was defeated conclusively in 1999 and several scholars argued that the party would never recover” \citep{plakoudas_insurgency_2018}. However, despite the fact that the PKK incurred heavy losses, these were mostly foot soldiers. The leadership structure was left intact, allowing the organization to regroup after only four years. 

\begin{figure}
  \caption{Preferential Attachment Over Time using EXIF Data}
      \begin{minipage}{\columnwidth}
  \includegraphics[width=\columnwidth]{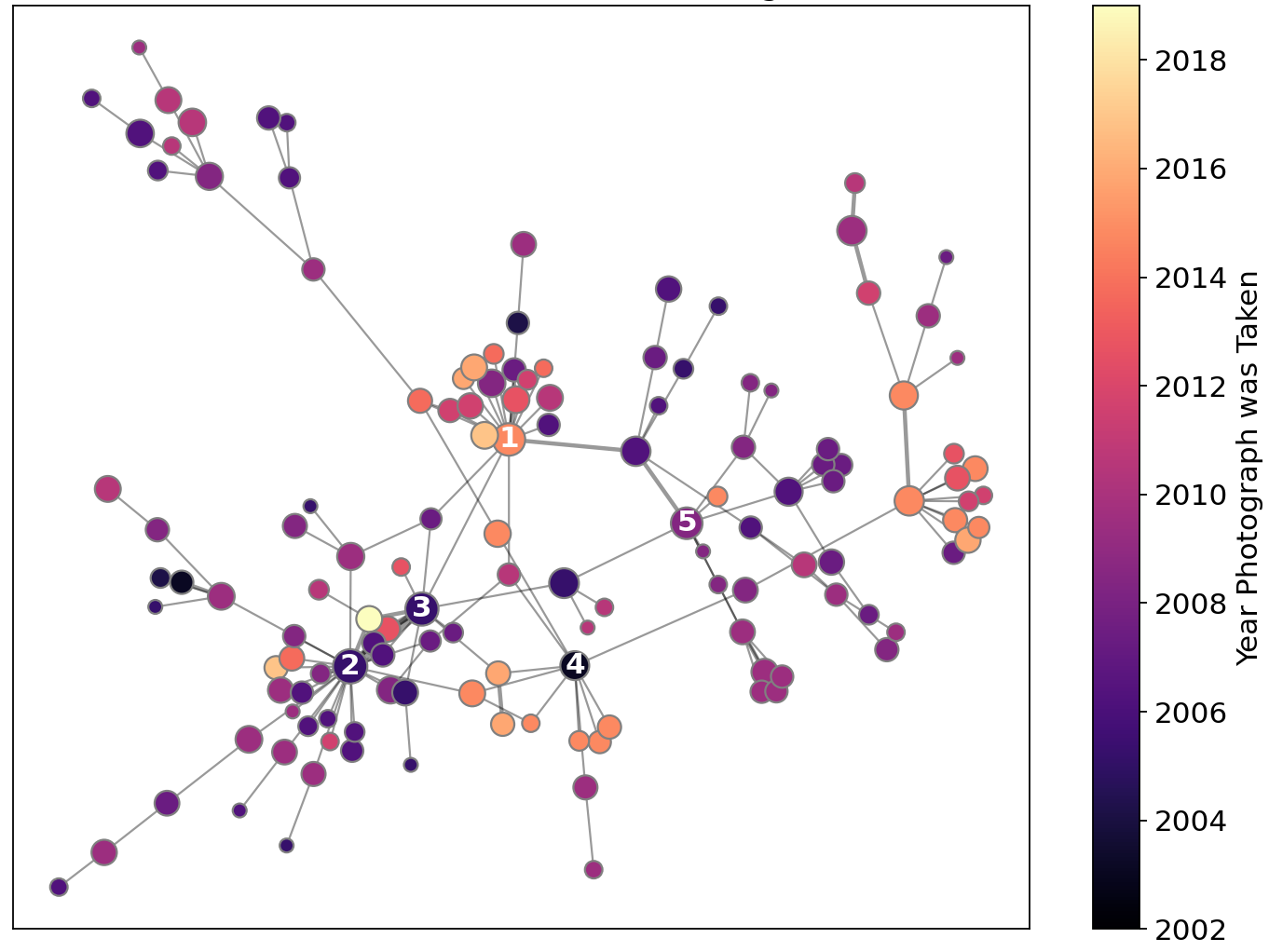}
{\scriptsize A network is constructed from metadata-containing images. Nodes are colored according to the oldest timestamp in that node's image cluster. The five most central nodes are numbered, and correspond to leaders. Newer nodes tend to cluster around older nodes.\par}
\end{minipage}     
\label{figure13}
\end{figure}

Following its military defeat in 2000 and re-emergence in 2004, the PKK grew rapidly. Figure \ref{figure13} utilizes image metadata to track this growth, showing the evolution of the network over time. Timestamps from photographs are extracted, and the earliest timestamp for each node is chosen thereby indicating an individual’s first known appearance in the field. Of the images that were sorted into clusters, 41\% (6373) contained machine-readable timestamps. This analysis should be taken as an approximation of the PKK’s regrowth rather than a precise model thereof. Many images in the sample do not contain readable timestamps, meaning that an individual may have been present in the PKK for far longer than the length of time suggested by the metadata; For example, the earliest available timestamp for Duran Kalkan is in 2014, despite his cluster containing several photos that appear far older. 

The top 5 most central nodes (in terms of betweenness) in this restricted sample are labeled, and once again correspond to the PKK leadership. Node 1 is Duran Kalkan, who despite the light color of his node has been in leadership roles within the PKK since the 1990s. The bulk of his neighbours appeared between 2008 and 2014. Nodes 2 and 3 are Bahoz Erdal and Kadir Celik, two senior figures in the military wing of the PKK, the HPG. They co-appear most frequently with individuals who first appear in the early 2000s, though their sustained presence in the field is indicated by co-appearances with much more recent recruits as well. Node 4 is Ali Haydar Kaytan, a co-founder of the PKK who largely served in non-combat roles. Interestingly, he appears almost exclusively with fairly recent recruits who first appear from 2014 onwards. Node 5 is Mehmet Gurhan, a military commander whose neighbours first appeared in the early 2000s. Despite incurring heavy losses throughout the 1990s, the PKK was able to regroup around a small number of core members. 

Figure \ref{figure13} shows some evidence of preferential attachment, whereby new nodes attach preferentially to already well-connected nodes. As newer, lighter nodes appear, they tend to attach to darker and more well-connected nodes. In other words, as newer recruits begin to appear in photographs, they tend to appear alongside older, more established members. Importantly, it seems that different elements of the leadership were important hubs at different periods in the PKK’s resurgence. This likely reflects the fact that “the PKK developed an advanced organizational structure, which included both military units and those elements responsible for recruitment activities, ideological training, propaganda efforts in order to increase awareness about the Kurdish question, and fundraising.” \citep[p. 728]{pusane_turkeys_2015}. Erdal, Celik, and Gurhan, members of the military leadership, were important hubs in the immediate aftermath of the PKK’s return to armed struggle in 2004. Political leaders, such as Kaytan, Kalkan, and Karaylian, act as important hubs for more recent recruits, particularly those appearing since 2014. The PKK’s ability to withstand generalized attacks (i.e. random node removal), as well as its regrowth around key members (preferential attachment) are characteristics associated with scale free networks. Thus, the structure of the PICAN aligns closely with the historical resilience of the PKK itself.

\subsection{Targeted Strikes}

Assuming the topology of a co-appearance network generated from militant photographs loosely approximates the general structural characteristics of the group itself, the former can be used to make general inferences about the future success of more recent counterinsurgency strategies. 

Seemingly in recognition of the failure of its previous counterinsurgency strategy, the Turkish government has embarked on a radically different approach in recent years. Particularly since 2016, Turkish forces have focused on targeting the leadership structure of the PKK through the use of airstrikes and raids both within and beyond Turkey’s borders \citep{tastekin_anatomy_2016}. In 2016, Turkish forces erroneously claimed to have assassinated Bahoz Erdal, the current leader of the PKK’s armed wing. The claim was spread largely by pro-government media, and indicated that he had survived at least two prior assassination attempts \citep{daily_sabah_pkk_2020}. Rather than attempting to inflict the largest possible number of casualties against the PKK by engaging in a conventional ground war, Turkish forces appear to be focusing on targeting key individuals within the organization.

In theory, a counterinsurgency strategy predicated on targeting the senior leadership of the PKK should be more disruptive to the network than one based on the removal of nodes at random; given the scale-free nature of the network, “disabling just a few critical nodes can result in a disconnected network especially for the smaller nodes” \citep{verma_economics_2020}. An empirical assessment of this claim in reference to the PKK fills an important gap in the applied literature on the social network analysis of insurgent groups: “claims that scale-free militant networks are resilient due to their robustness against random attacks on nodes have not adequately reckoned with the countervailing effect that high-degree nodes are more visible” \citep[p. 233]{zech_social_2016}.

The selection of targets in this strategy is likely largely opportunistic; previous operations have targeted PKK leaders on the basis of available intelligence, which often results from chance encounters. In May 2021, the Turkish President announced the assassination of Sofi Nurettin, a PKK military officer responsible for the group’s presence in Syria \citep{sabah_turkey_2021}. The operation was carried out by an intelligence unit formed specifically to target Nurettin, which ultimately located him by tracking his family \citep{sabah_turkey_2021}. Nurettin survived an assassination attempt in 2016, suggesting that Turkish forces had been targeting him for at least 5 years \citep{tastekin_anatomy_2016}. Beyond entailing a significant investment in terms of military and intelligence resources, the operation’s success hinged significantly on luck. Thus, the Turkish government’s new military strategy involves targeting PKK leadership, but only managing to strike opportunistically. 

\begin{figure}
  \caption{PICAN Robustness to Opportunistic Central Node Removal}
      \begin{minipage}{\columnwidth}
  \includegraphics[width=\columnwidth]{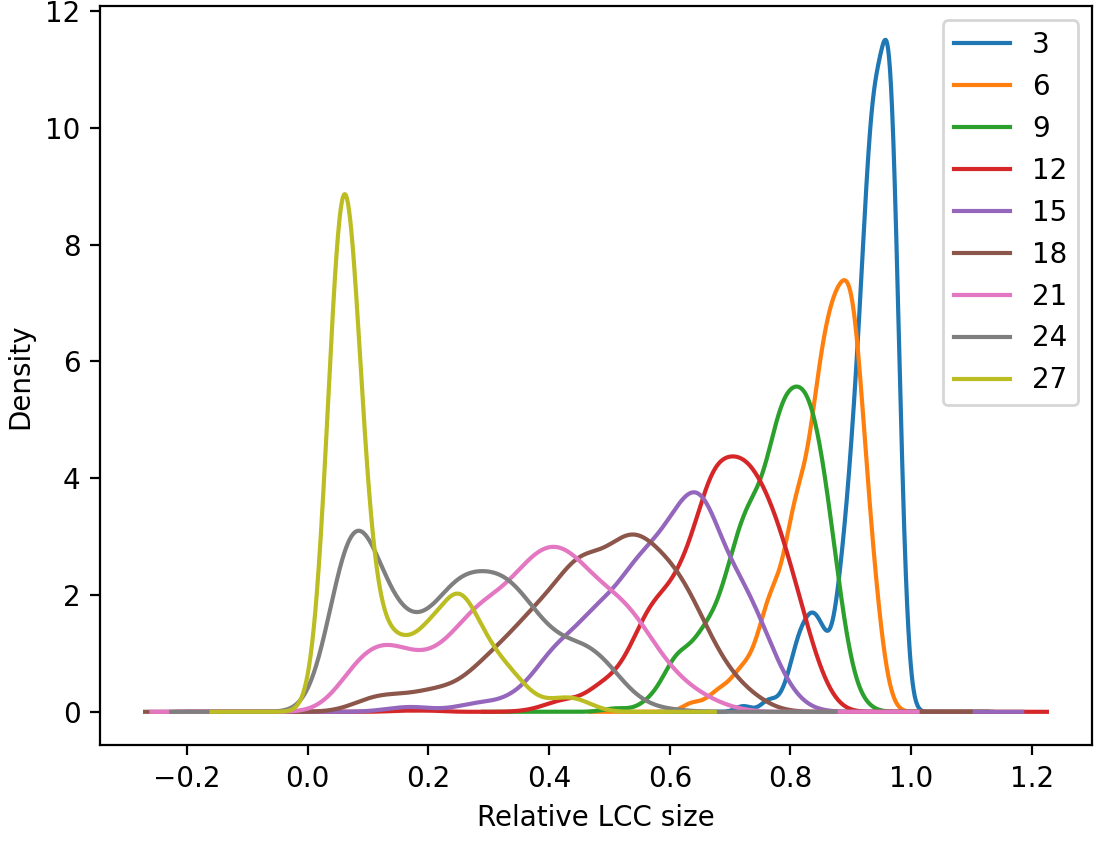}
{\scriptsize Distributions reflect the effect of removing a random set of nodes from the top 30 most central nodes on the relative size of the LCC. Colors indicate the number of nodes removed, a procedure which is repeated 1000 times. High levels of connectivity are maintained even when the majority of the top 30 nodes are removed.
\par}
\end{minipage}     
\label{figure14}
\end{figure}

The distributions shown in Figure \ref{figure14} approximate the effects of such a strategy on the LCC by randomly removing a given number of nodes T from the set of 30 most central nodes. Each level of T roughly corresponds to the aggressivity of a counterinsurgency strategy; the removal of only 3 random nodes out of the top 30 most central nodes would correspond to a relatively weak counterinsurgency approach, while the removal of 27 out of 30 corresponds to an extremely aggressive strategy. In practice, the value of T is likely to be determined by available opportunities. 1000 random sets of nodes are removed at each level, and the resulting kernel density estimates of the effect on LCC size are reported. 

If only three out of the top 30 most central nodes are removed (the scenario indicated by the distribution in blue), there is virtually no effect on network connectivity. As more nodes are removed, connectivity decreases but the variance of the distributions increases substantially. Thus, the extent to which the removal of, for example, 15 nodes decreases the size of the LCC depends heavily on which 15 nodes are removed. Bimodality in some of these distributions likely indicates splintering brought about by either the removal or survival of the three most central nodes. When 27 of the top 30 most central nodes are removed, there is a strong likelihood of the relative LCC size dropping to near zero. However, the smaller second peak on the right of the distribution indicates that the LCC can retain roughly 20\%-- and in rare cases over 40\%-- of its original size, depending on which three nodes out of the top 30 survive. Similarly, though the removal of only 3 of the top 30 nodes overwhelmingly results in the LCC remaining unchanged, a much smaller peak to the left of the distribution indicates potential reductions in LCC size of up to 20\%, presumably in the unlikely event that the three most central nodes are removed.
\begin{figure*}[h]
  \caption{Comparison between PICAN and a Random Graph}
\begin{minipage}{\textwidth}
  \includegraphics[width=\textwidth]{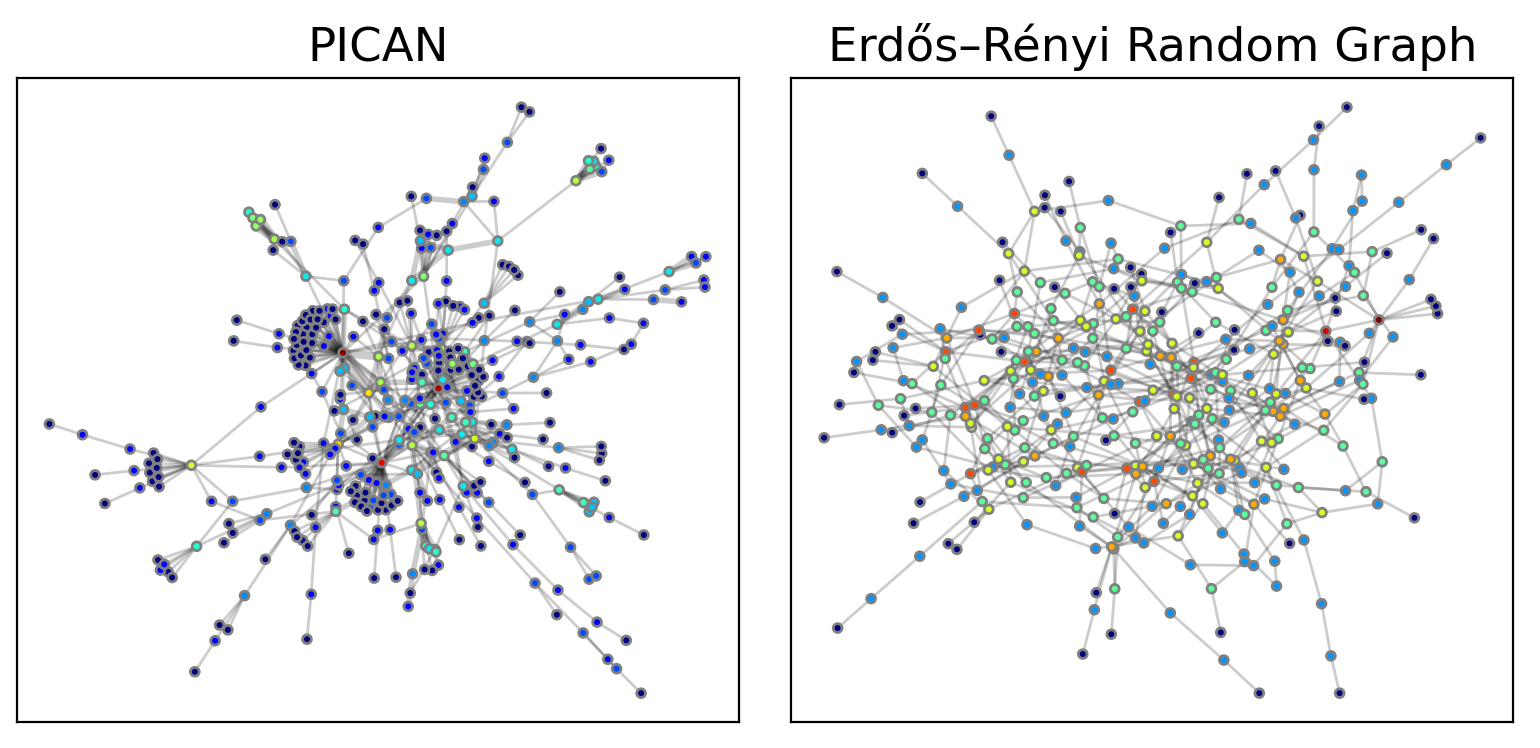}
{\scriptsize Compared to a random graph with the same number of nodes and edges, the PICAN features a number of high-degree hub nodes (shown in red). There also appears to be a higher degree of clustering and interconnectivity in the PICAN. 
\par}
\end{minipage}  
    \label{figure15}
\end{figure*}
Though the theoretical literature on scale-free networks suggests that these are highly vulnerable to the removal of hub nodes, the PICAN appears to be surprisingly robust to such a strategy. Even when a random set of 15 of the top 30 most central nodes is removed, the network tends to maintain well over 60\% of its connectivity. Thus, while the top three most central nodes in the PICAN stand out in terms of their centrality, a significant amount of the network’s connectivity is distributed across a larger set of important figures. Indeed, the FARC-- a similarly hierarchical paramilitary organization-- was able to withstand the assassination of 53 of its leaders and three members of its Secretariat \citep{segura_made_2017}.

This type of resilience is characteristic of “small world” networks. This class of networks overlaps with (and shares many of the properties of) scale-free networks; they are characterized by the presence of “hub” nodes, and are thus generally robust to random node removal. Small world networks are further characterized by a short average path length; compared to a random graph with the same number of nodes and edges, the presence of well-connected hubs enables shorter paths between sets of nodes. Figure \ref{figure15} compares the Largest Connected Component of the PICAN with an Erdős–Rényi random graph with the same number of nodes and edges, the former displaying the aforementioned hub-and-spoke configuration.

In the PICAN, the average shortest path between nodes is 4.9; the random graph has an average shortest path length of 6.4. This is largely due to the presence of high-degree hub nodes that are clearly visible in the PICAN but absent in the random graph. The nodes are colored by degree, with the random graph exhibiting a much more uniform degree distribution. Importantly, the PICAN not only contains well-connected hub nodes, but general areas of high connectivity. This type of interconnectivity is also typical of small world networks, which exhibit a higher degree of clustering-- the extent to which a node’s neighbours are also interconnected. A quantitative measure of the extent of clustering in a graph is given by the \cite{watts_collective_1998} clustering coefficient, defined as:

$${c}_{i}^{ws}=\frac{{2E}_{i}}{{k}_{i}({k}_{i}-1)}$$

Where $E_{i}$ is the number of edges between neighbours of node i, and K is the degree of the node. The more interconnected a node’s neighbours are, the higher the clustering coefficient, and the more robust that network will be to the removal of the node. The average clustering coefficient of the PICAN is 0.0574 compared to just 0.0014 for the random graph, meaning that the former is more densely interconnected and therefore robust to the removal of even high-degree nodes.

\cite{humphries_network_2008} propose a formal test for small-world-ness (3) based on the premises that, compared to a random graph, (1) a small-world network will have a shorter average path length L and (2) a higher degree of clustering C:

\begin{equation} 
{\lambda}_{g}=\frac{{L}_{g}}{{L}_{rand}}
\end{equation}

\begin{equation} 
{\gamma}_{g}^{ws}=\frac{{C}_{g}^{ws}}{{C}_{rand}^{ws}}
\end{equation}

\begin{equation} 
{S}^{ws}=\frac{{\gamma}_{g}^{ws}}{{\lambda}_{g}}
\end{equation}

If the above conditions are met, then S>1 and the network may be considered a small world. The value of S for the PICAN is 2.69, suggesting that the organizational structure of the PKK resembles that of a small world network. 

In a study of jihadist insurgent groups, \cite[p. 140]{sageman_understanding_2011} notes that “Unlike a hierarchical network that can be eliminated through decapitation of its leadership, a small-world network resists fragmentation because of its dense interconnectivity”. Jemaah Islamiyah, a Southeast Asian jihadist group, was effectively incapacitated following the arrest of its senior leadership; the group’s extremely hierarchical nature left local cells unable to function without direct orders from above \cite[p. 141]{sageman_understanding_2011}. Despite the presence of a few key hub nodes in the PICAN, there also exists a large pool of highly interconnected nodes representing the second and third tiers of the organization’s leadership. This affords the PKK a high level of robustness to the targeting of even its most central figures. Furthermore, the node removal simulations carried out in Figure \ref{figure14}
 are carried out on a static graph. As shown in the analysis of the PKK’s regrowth shown in Figure \ref{figure13}, the network’s evolution over time is characterized by preferential attachment to hub nodes. Not only would the survival of even a small number of these nodes allow for the regrowth of the group; “new hubs will take the role of the eliminated ones and restore the network’s ability to function.” \cite[p. 140]{sageman_understanding_2011}.

These results suggests that the robustness of a co-appearance network generated solely from militant photographs to certain node removal scenarios broadly aligns with the true network's historical resilience to analogous counterinsurgency strategies. The scale-free structure of the PICAN renders random node removal strategies ineffective, as connectivity is maintained by the survival of a relatively small number of high-degree nodes. This is congruous with the historical record: “Considering the fact that the PKK has developed such a complex system of networks and institutions, as well as various sources of funding both at the domestic and international levels, it was no surprise that Turkey’s ability in weakening the PKK militarily in the late 1990s did not bring an end to the PKK insurgency” \cite[p. 730]{pusane_turkeys_2015}. Assuming even a tenuous relationship between the PICAN and the actual structure of the PKK, the image co-appearance network is able to provide contextual information relevant to an assessment of the future success of the counterinsurgency strategy currently being undertaken by the Turkish government. As a small world network, the PICAN tends to survive the removal of the majority of its most central nodes due to dense interconnectivity among them. Indeed, the preceding paragraph suggests that the PICAN probably underestimates the extent of this interconnectivity. The corresponding real-world conclusion is that the degree of hierarchy within the PKK is not so extreme as to enable the group to be incapacitated by the removal of a few key leaders. Thus, even a strategy of targeted strikes is unlikely to be effective against the PKK, especially when one considers the high cost and low success rate of such efforts.

\section{Conclusion}

The analytical approach elaborated in this study makes three distinct contributions to the literature on the application of Social Network Analysis to the study of insurgent groups. Firstly, it leverages an abundant but underutilized source of primary data generated by militants themselves. While the main challenge in the study of dark networks is the lack of data thereon, a defining trend in the evolution of modern insurgency has been the use of online media, particularly online visual propaganda \citep{dauber_youtube_2009}. Social network analysts have recognized the rich relational information contained in image co-appearances as a valuable tool for understanding social structures \citep{lewis_tastes_2008, berry_friends_2006, golder_measuring_2008}. By automating the process of generating an image co-appearance network through facial recognition and unsupervised clustering, the present approach is able to harness the vast quantities of unstructured image data generated by insurgent groups. The distribution of this methodology as an open source python package further facilitates its use by researchers of dark networks. 

The ability to process large quantities of primary data enables a second significant contribution to the existing literature, which relies almost exclusively on secondary sources. Social networks generated through media reports, legal filings, and public statements require a significant amount of pre-existing knowledge about the group in question, and inherently omit all but the highest-ranking figures. Though selection bias in insurgent photographs is still a problem, a foot soldier is probably more likely to appear in the background of a group photograph than to be named in a newspaper. Image co-appearance networks are thus likely to be far more complete, and can be created without additional knowledge about the group in question. 

Finally, the use of co-appearances imposes a level of consistency in the ordering of the network. In previous studies of insurgent groups, ties between nodes have encompassed everything from marriage to casual acquaintance \citep{gill_lethal_2014}, from cohabitation to “weak contact” \citep{koschade_social_2007}. To be sure, co-appearance in militant photographs encompasses a wide range of possible social interactions, from friends sharing a meal to a commander posing with her subordinates. If two people are co-present in a photograph, the nature and social significance of the link between them is unknown. However, if the same person repeatedly appears with others in a large number of photographs, there is probably an overarching reason. An image co-appearance network mirrors the social forces that govern when and where pictures are taken. In the context of an insurgency where virtually all social interaction is mediated by the group’s structure, an image co-appearance network will generally reflect these ordering principles, including divisions between military units, ranks, and political factions: foot soldiers will co-appear because they’re in the same platoon; commanders officiating training academy graduations will appear once with each graduand and frequently with each other; leaders who have fallen out are unlikely to pose together. The network will also reflect performative aspects of the group’s self-perception, emphasizing actors and narratives that align with the image militants wish to present to the world, the precise nature of which depends on the ideological characteristics of the group in question. 

Thus, the use of co-appearances in militant photographs enables detailed insights into the social, political, and organizational forces that structure insurgent groups. The three analytical sections in this study demonstrate the types of analyses that are possible with a co-appearance network generated by militant photographs, using a large quantity of obituary images taken by the PKK. 

Section 5 demonstrates that a qualitative analysis of the nature of an individual’s co-appearances can yield information on functional and factional divisions within a rebel group. In the PKK, military commanders have high betweenness centrality due to frequent co-appearances with subordinates, who are themselves generally poorly connected nodes. Political leaders spend less time with the rank and file, and more time with each other, leading to higher eigenvector centrality. The nature of co-appearances even captures the effects of Cemil Bayik’s marginalization following a failed leadership struggle: despite being a founding member of the PKK, he has relatively few co-appearances and appears in no photographs with the top leaders of the organization.

Section 6 demonstrates a consistent relationship between an individual’s rank and their centrality in the co-appearance network. This is achieved by matching over 100 nodes in the PICAN with individuals appearing on lists of wanted persons maintained by the Turkish government, with the value of the reward placed on them acting as a proxy for rank. Linear models indicate that over half of the variation in the reward offered for an individual’s capture can be explained purely by the number of times they appear with others in photographs. Exponential Random Graph Models show strong positive associations between reward value and the likelihood of edge formation, suggesting that higher ranking individuals are more well-connected in the PICAN. In both sets of models, results are robust to the exclusion of the top leadership from the analysis. 

Section 7 shows that the co-appearance network may even approximate the general structural characteristics of the insurgent group. A close reading of the topology of the PICAN in reference to qualitative work on the PKK’s history finds agreement between the PKK’s resilience to generalized attacks and the PICAN’s structural robustness to an analogous node removal strategy. If the PICAN can be assumed to roughly approximate the structure of the PKK, then the Turkish military’s current approach of targeted strikes against leadership is also unlikely to be effective; a simulation of targeted node removal shows that the network’s connectivity is maintained by a large and densely interconnected pool of high ranking individuals waiting in the wings.

While insights from these three sections are specific to the nature of the image dataset and the insurgent group in question, images gathered from other sources or rebel groups would manifest their own idiosyncrasies and highlight different social and organizational processes. For example, an egocentric network of a single insurgent cell or unit could be constructed from an individual’s posts on social media, with temporal and geographical nodal attributes derived from metadata. Footage from the commission of war crimes-- disturbingly rife online-- could be used to understand whether incidents are occurring at random, or whether the same individuals systematically co-appear. Images collected from multiple groups could reveal the existence and nature of links between members of both organizations. Though dark groups are exceptionally difficult to study, the extraction of relational information from images in the form of co-appearances between individuals can yield detailed information on their social structure. 

\newpage

\onecolumn
\section*{Appendix} \label{sec:tab}
\setcounter{table}{0}
\renewcommand{\thetable}{A\arabic{table}}
\setcounter{figure}{0}
\renewcommand{\thefigure}{A\arabic{figure}}

\begin{figure}[hp]
  \caption{Mosaic of Input Images scraped from online PKK obituaries}
      \begin{minipage}{\textwidth}
  \includegraphics[width=\textwidth]{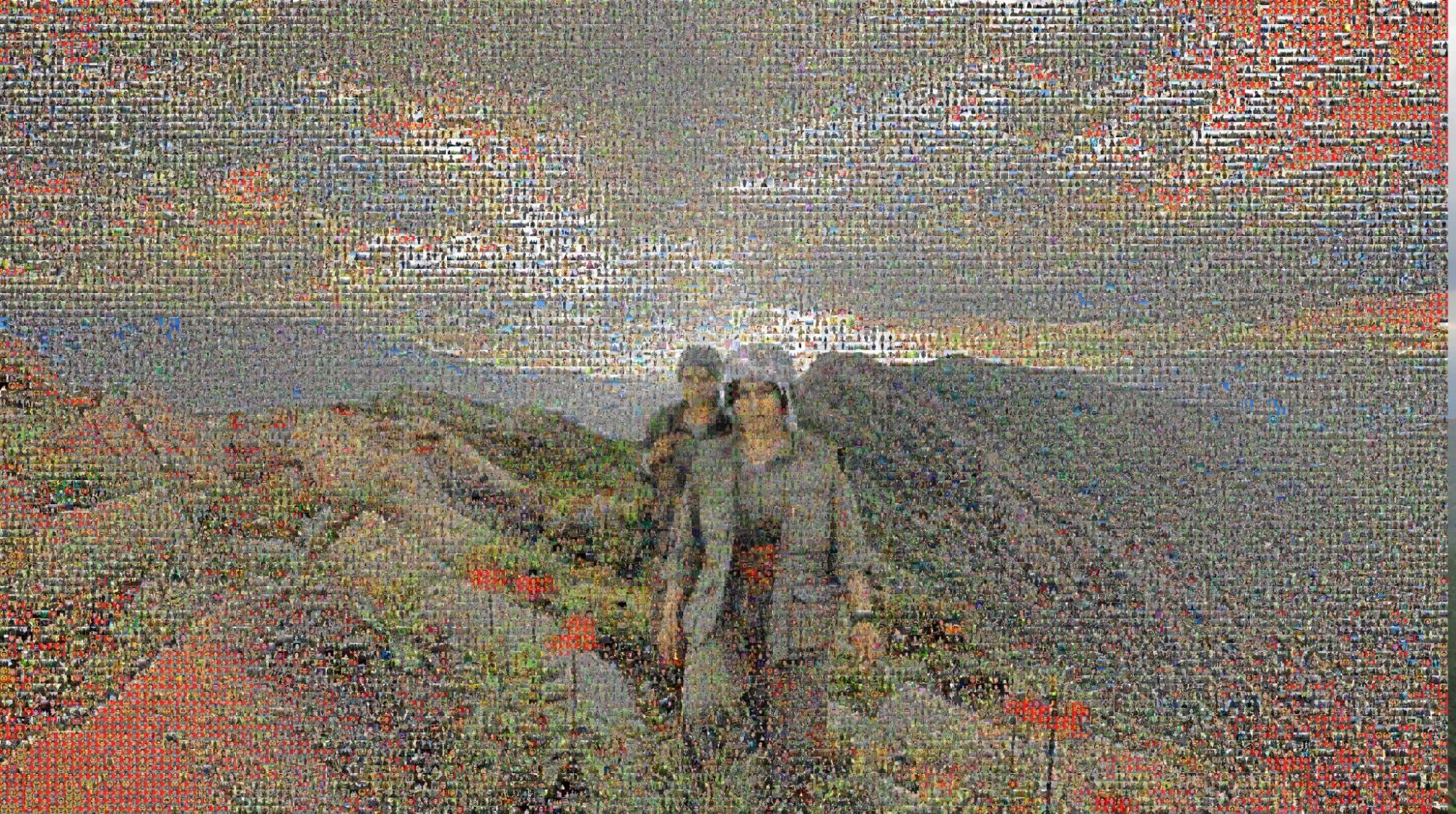}
{\footnotesize 
This image mosaic was constructed from roughly 20,000 publicly available images posted by the PKK to their online obituaries website, \href{https://hpgsehit.com}{hpgsehit.com}. A version of this figure with an interactive zoom functionality to view individual pictures can be accessed \href{https://oballinger.github.io/PICAN-data/}{here}.

\par}
\end{minipage}     
\label{A1}
\end{figure}

\begin{figure}[hp]
  \caption{Sample Obituary}
      \begin{minipage}{\textwidth}
  \includegraphics[width=\textwidth]{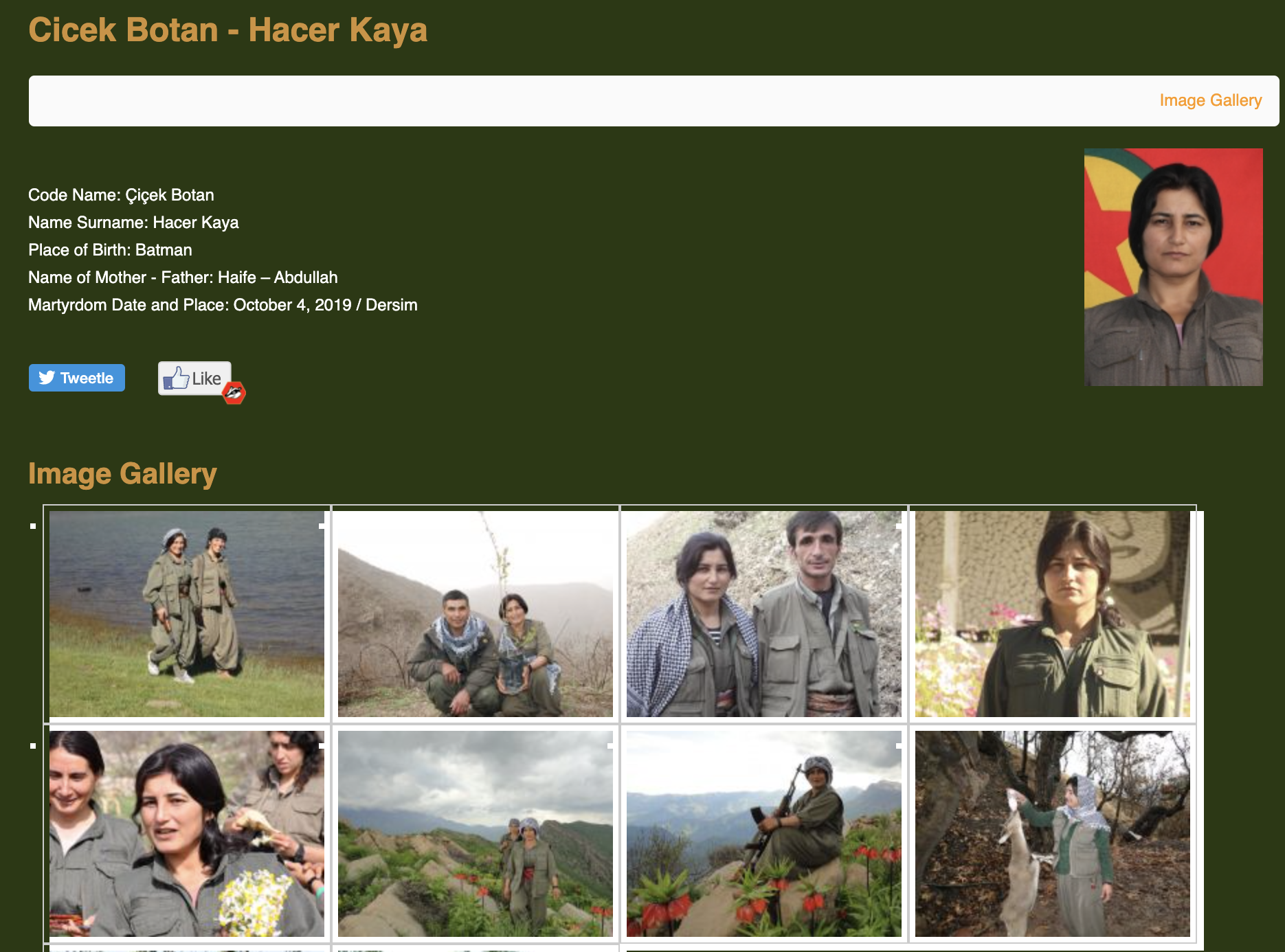}
{\footnotesize The screenshot above was taken from an entry on the PKK's \href{hpgsehit.com}{obituary website}. The obituary includes a high defitinition portrait in the top right, some basic biographical details on the left, and a set of images of the fallen militant. 
\par}
\end{minipage}     
\label{A2}
\end{figure}

\begin{figure}[hp]
  \caption{Representations of Martyred PKK Leaders}
      \begin{minipage}{\textwidth}
  \includegraphics[width=\textwidth]{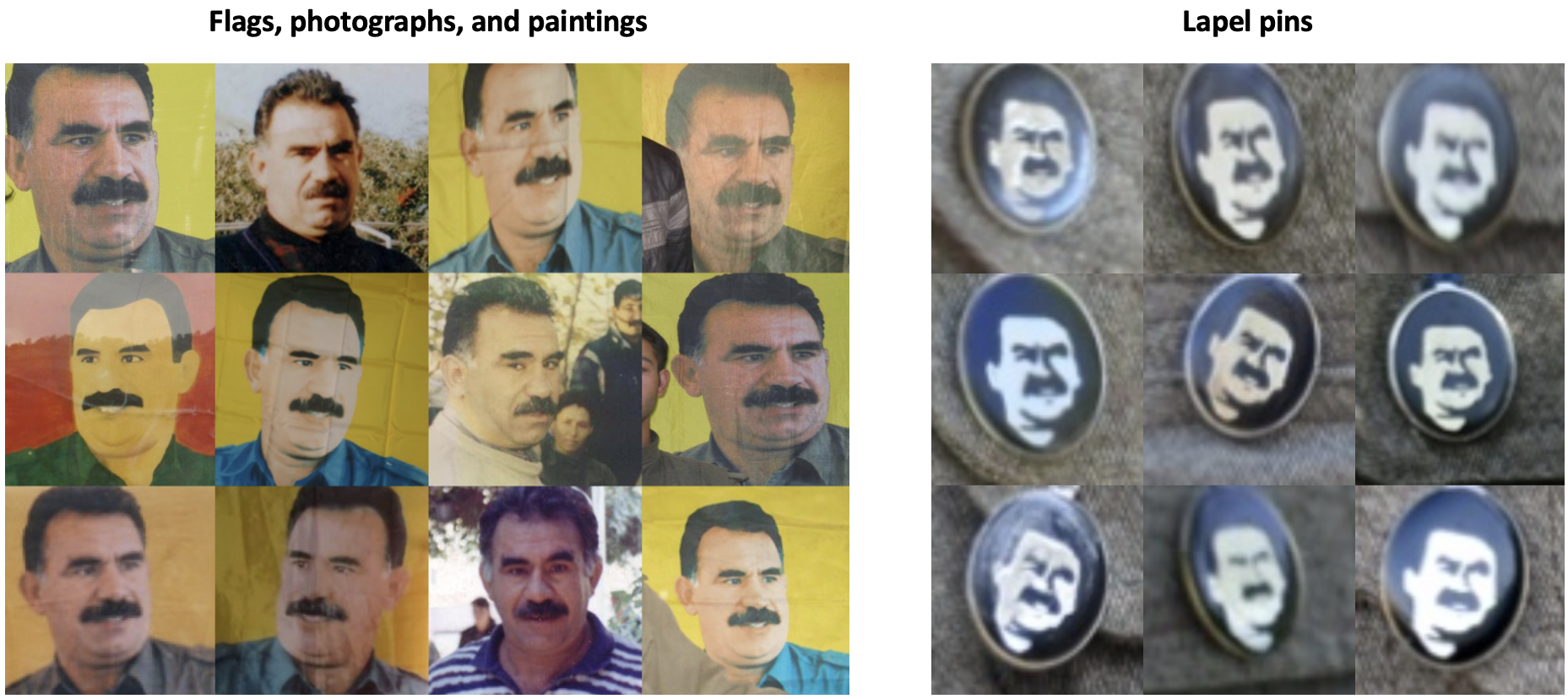}
{\footnotesize The images above are samples from facial clusters containing representations of two PKK leaders who are no longer in the field, but whose likeness is still present in the photographs. On the right, the likeness of Mahzum Korkmaz is depicted on lapel pins which appear in 21 photographs. He was the PKK's first military commander, and despite being killed in 1986, remains an important figure in the PKK. The images on the left were taken from the cluster generated for PKK founder Abdullah Öcalan, who was arrested and imprisoned in 1999. His likeness is depicted on flags and murals that appear in the background of 107 photographs. Interestingly, despite not being in the field for the past 23 years, his node in the PICAN is the most central. Both of these phenomena reinforce the notion that there is a relationship between an individual's importance to the PKK and their embedding in the image co-appearance network. However, for the sake of conceptual clarity in the definition of edges in the PICAN, these nodes are removed from the network. 
\par}
\end{minipage}     
\label{A3}
\end{figure}

\begin{figure}[hp]
  \caption{U.S. State Department Wanted Poster for PKK Leadership}
      \begin{minipage}{\textwidth}
      
  \includegraphics[width=\textwidth]{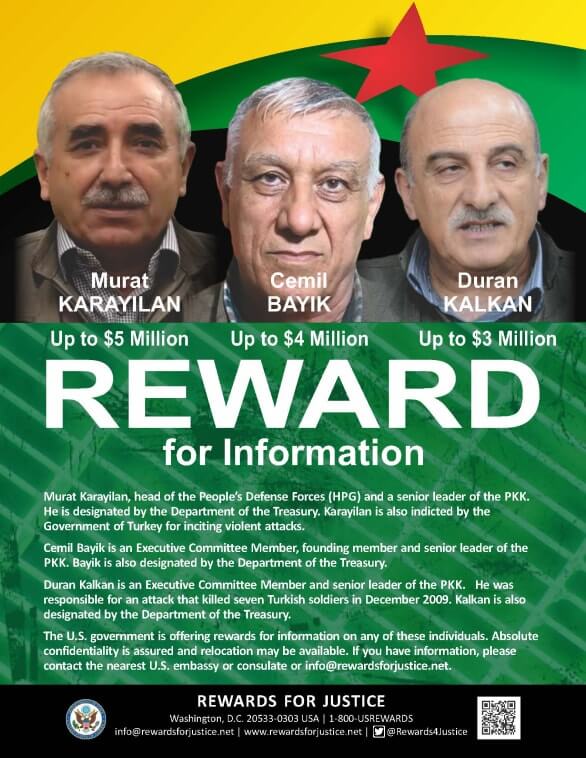}
{\footnotesize 
This U.S. State Department wanted poster details rewards for information leading to the capture of PKK leaders Murat Karayilan, Cemil Bayik, and Duran Kalkan. All three individuals occupy highly central positions in the PKK Image Co-Appearance Network; They are respectively the 2nd, 8th, and 3rd most central nodes in the PICAN in terms of betweenness centrality. 

\par}
\end{minipage}     
\label{A4}
\end{figure}

\begin{figure}[hp]
  \caption{Central University Research Ethics Committee Approval Letter}
  \includegraphics[width=\textwidth]{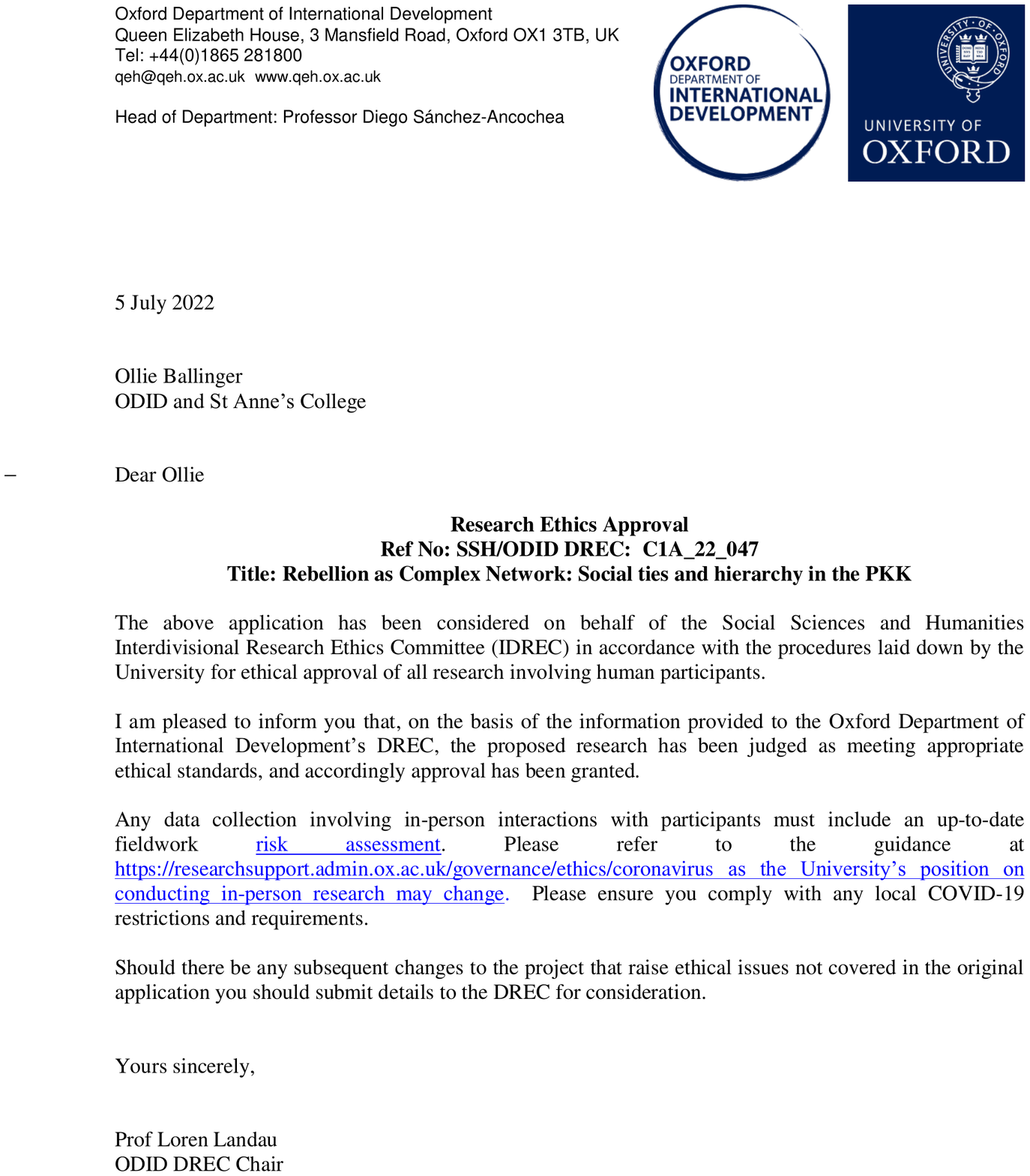}
\label{ch3_curec}
\end{figure}

\newpage
\include{face_network}

\twocolumn
\newpage
\bibliographystyle{unsrtnat}
\bibliography{references} 

\end{document}